%% file: main.tex
\documentclass[lettersize,journal]{IEEEtran}
\usepackage{amsmath,amsfonts,amssymb}
\usepackage{algorithmic}
\usepackage{array}
\usepackage[caption=false,font=normalsize,labelfont=sf,textfont=sf]{subfig}
\usepackage{textcomp}
\usepackage{stfloats}
\usepackage{graphicx}
\usepackage{url}
\usepackage{verbatim}
\usepackage{multirow}
\usepackage{caption}
\usepackage{diagbox}
\def\BibTeX{{\rm B\kern-.05em{\sc i\kern-.025em b}\kern-.08em
    T\kern-.1667em\lower.7ex\hbox{E}\kern-.125emX}}
\usepackage{xcolor}

\usepackage{balance}
\usepackage{amsthm}
\newtheorem{Theorem}{Theorem}
\newtheorem{Lemma}{Lemma}

\newtheorem{Proposition}{Proposition}
\newtheorem{Definition}{Definition}

\newtheorem{Remark}{Remark}

\begin{document}
\title{Multiplexed Streaming Codes for Messages with Different Decoding Delays in Channel with Burst and Random Erasures}
\author{Dingli Yuan, Zhiquan Tan, Zhongyi Huang
\thanks{This work was partially supported by the NSFC Project No. 12025104.}}

\maketitle
\begin{abstract}
In a real-time transmission scenario, messages are transmitted through a channel that is subject to packet loss. The destination must recover the messages within the required deadline. In this paper, we consider a setup where two different types of messages with distinct decoding deadlines are transmitted through a channel model that introduces either one burst erasure of length at most $B$, or $N$ random erasures in any fixed-sized sliding window. The message with a short decoding deadline $T_{\mathrm{u}}$ is referred to as an urgent message, while the other one with a decoding deadline $T_{\mathrm{v}}$ ($T_{\mathrm{v}} > T_{\mathrm{u}}$) is referred to as a less urgent message.

We consider the scenario where $T_{\mathrm{v}} > T_{\mathrm{u}} + B$ and propose a non-trivial achievable region $\mathcal{R}$ for the aforementioned channel model. We propose a novel merging approach to encode two message streams of different urgency levels into a single flow and present explicit constructions for encoding, contributing to the establishment of the achievability of region $\mathcal{R}$. Our comprehensive analysis demonstrates that this region encompasses the rate pairs of existing encoding schemes and coincides with the capacity region in burst channel scenarios. Lastly, we investigate the property of the achievable region $\mathcal{R}$, proving that it is the largest one obtained from all the rate pairs under the merging method.

\end{abstract}
\begin{IEEEkeywords}
Forward error correction (FEC), sliding window, maximum distance separable (MDS) codes, multiplexing, streaming codes.
\end{IEEEkeywords}
\section{Introduction}
\IEEEPARstart{I}n the last decade, low-latency real-time communications and high-definition video applications have become increasingly important \cite{cisco2018}. There are numerous practical applications, including Augmented Reality (AR)/Virtual Reality (VR), audio and video conferencing, remote healthcare, industrial automation, smart grid, and many other scenarios \cite{feng2021}. The challenge of implementing delay-constrained communications is further exacerbated by noise, interference, fading, routing, mobility, and reliability requirements. To reduce packet loss in communication, we must introduce strategies to mitigate it.

To control packet loss, data retransmission technologies such as the Automatic Repeat Request Protocol (ARQ) \cite{lin1984repeat} or Forward Error Correction (FEC) \cite{badr2017survey} are usually deployed. Although ARQ generally introduces a minimal amount of redundancy, each retransmission often increases the round-trip delay, which may exceed the delay requirement for long-distance communication. Moreover, ARQ results in a more complex protocol because the transmitter requires confirmation from the receiver. If a message is received but its acknowledgment is lost, the sender will have to resend it, wasting time and bandwidth \cite{lin1984repeat}. On the other hand, FEC techniques, by introducing redundancy, provide an efficient alternative solution, where the receiver can possibly recover the lost packets independently through the redundancy. The challenges lie in constructing streaming codes so that the messages can be decoded within the deadline and determining how much redundancy should be introduced for a specific channel model with a given decoding deadline.

Taking Maximum Distance Separable (MDS) codes \cite{macwilliams1977theoreyECC} as an example, such as Reed-Solomon (RS) codes \cite{reed1960}, they have good recovery for burst erasure but may cause a significant delay. In \cite{martinian2004burst}, Martinian and Sundberg characterize the capacity (maximum code rate) of channels with a burst erasure of length at most $B$ and provide a construction that meets the capacity $C(T, B) = \frac{T}{T + B}$, where $T$ is the decoding deadline. Furthermore, in \cite{badr2016layered}, a channel model with both burst erasure and random erasures, i.e., \textit{sliding window model} is considered. It is assumed that for any window of length $W$, the channel can introduce one of the following two packet loss patterns: (1) a burst erasure of length at most $B$, (2) at most $N$ random erasures. The capacity under this setup is shown to be:
\begin{equation}\label{Eqn_capa_MS}
C(T,B,N) = \frac{T - N + 1}{T - N + 1 + B},
\end{equation}
where $W\geq T+1$.
Since then, a few constructions of streaming codes have been proposed to achieve the above capacity \cite{krishnan2020rate-optimal,domanovitz2021anexplicit}.

The aforementioned works only consider a single stream of data. However, the study of a single stream of streaming code may not be enough for practical applications. A typical multiplexed coding setup exists where information of varying importance requires differing levels of error protection, known as unequal error protection (UEP) \cite{masnick1967UEPlinear,haghifam2018streamingUEPburst}. Unlike standard error protection strategies, the importance of information is also measured by varying delay constraints.  For instance, internet applications often involve multiple types of data streams, including video, audio, text, etc., which have different decoding deadlines. For example, in remote teaching, one stream could comprise the audio/video content and will require a stringent delay constraint, while the other stream comprising presentation slides could have a more relaxed delay constraint. Another example is a novel network protocol named Quick UDP Internet Connection (QUIC) which is built on UDP \cite{dey2022QUIC}. It supports multiplexing and utilizes a XOR-based FEC method to recover dropped packets without waiting for retransmission. This illustrates the growing need to study multiplexed coding in the context of different delay constraints.

In the multiplexed streaming framework, we examine a scenario where two types of messages must be transmitted, each with a unique decoding deadline denoted as \( T_{\mathrm{u}} \) and \( T_{\mathrm{v}} \). Specifically, messages with a more stringent decoding deadline \( T_{\mathrm{u}} \) are designated as \textit{urgent messages}. Conversely, those messages with a more lenient decoding deadline \( T_{\mathrm{v}} \) are labeled as \textit{less urgent messages}. Both types of messages are co-encoded and transmitted within a single channel-packet stream.

In the literature, reference \cite{badr2018multi} provides a comprehensive study of the capacity region, represented as the convex hull of all feasible rate pairs \( (R_{\mathrm{v}}, R_{\mathrm{u}}) \), specifically for channels subject to burst erasures with lengths bounded by \( B \).  Additionally, a systematic streaming code that achieves the non-trivial point of the capacity region was introduced specifically for scenarios where $T_{\mathrm{u}} < T_{\mathrm{v}} < T_{\mathrm{u}} + B$. Building upon this foundational work, reference \cite{fong2020opmulti} investigates the complementary scenario where \( T_{\mathrm{u}} + B < T_{\mathrm{v}} \). In this extended setting, the paper thoroughly characterizes and delineates the capacity region. 

In this paper, we focus on the multiplexed streaming model for channels involving burst or random erasures. Specifically, we consider the transmission of two distinct types of messages: urgent messages with a decoding deadline of \( T_{\mathrm{u}} \), and less urgent messages with a decoding deadline of \( T_{\mathrm{v}} \). We define that within any window of size \( W \), messages can be successfully decoded within their respective deadlines if either a burst erasure of length no greater than \( B \) occurs, or no more than \( N \) random erasures occur. For the specialized case where \( T_{\mathrm{v}} > T_{\mathrm{u}} + B > T_{\mathrm{u}} > B \), we propose a non-trivial achievable region $\mathcal{R}$, which is strictly larger than the trivial one \(\mathcal{B}\), defined by the two extreme points \((C(T_{\mathrm{v}},B,N),0)\) and \((0,C(T_{\mathrm{u}},B,N))\). We start deriving a novel merging methodology to prove the achievability of $\mathcal{R}$. This merging method is based on the rate-optimal streaming codes for a single flow, as discussed in \cite{domanovitz2021anexplicit}.
The key steps in our approach are:

\begin{itemize}
    \item Utilize rate-optimal block codes from \cite{domanovitz2021anexplicit}, configured with appropriate parameters, to encode both urgent and less urgent messages, resulting in two separate encoded packages.
    \item Merge these two packages to minimize redundancy.
\end{itemize}

Furthermore, we propose explicit constructions for encoding the two streams of messages over a field size that scales quadratically with the delays. With these constructions, we can obtain a non-trivial corner point in the rate pair region, subsequently derive the achievable region $\mathcal{R}$.  Comparative analysis demonstrates that for any choices of \(B\) and \(N\), the rate pair region of the existing encoding schemes are contained within the achievable region \(\mathcal{R}\). Notably, in the special case where \(B > N = 1\), which corresponds to the burst channel, \(\mathcal{R}\) coincides with the capacity region as stated in \cite{fong2020opmulti}.
Additionally, we explore the property of the achievable region $\mathcal{R}$ and demonstrate that it represents the largest possible region derived from all the rate pairs using the merging method.

The paper is structured as follows: Section II provides the foundational model and necessary preliminaries that will be referenced throughout the paper. Section III illustrates the main results of our work. Section IV details our novel merging method, and additionally presents explicit constructions for encoding. Section V provides a detailed proof demonstrating the achievability of the proposed constructions. Section VI details the property of region $\mathcal{R}$. Section VII concludes the article.

\subsection{Notations}
The sets of natural numbers, integers, non-negative integers, and non-negative real numbers are denoted by $\mathbb{N}, \mathbb{Z}, \mathbb{Z}_{+}$ and $\mathbb{R}_{+}$ respectively.
All the elements of any matrix considered in this paper are taken from a common finite field $\mathbb{F}$, where 0 and 1 denote the additive identity and the multiplicative identity respectively.
The set of $k$-dimensional row vectors over $\mathbb{F}$ is denoted by $\mathbb{F}^{k}$, and the set of $k \times n$ matrices over $\mathbb{F}$ is denoted by $\mathbb{F}^{k \times n}$. 
A row vector in $\mathbb{F}^{k}$ is denoted by $\mathbf{a} \triangleq[a_{0}, a_{1},\dots, a_{k-1}]$, where $a_{\ell}$ denotes the $(\ell+1)$-th element of $\mathbf{a}$. 
The $k$-dimensional identity matrix is denoted by  $\mathbf{I}_{k}$. 
The interval $I \triangleq [a:b]$ represents the set of all integers from $a$ to $b$, inclusive. 
The interval $I_1\backslash I_2$ represents the set of all elements that are in interval $I_1$ but not in interval $I_2$. 
A column vector $G_{[i]}$ represents the $i$-th column of matrix $G$. 
A sub-matrix $G_{[a:b]}$ refers to the sub-matrix of $G$, spanning from the $a$-th column to the $b$-th column, inclusive.
A unit vector $\eta_j $ denotes the unit vector with its $j$-th component set to 1, and all other components set to 0.

\section{Background knowledge}
\subsection{Preliminaries}

\begin{Definition} \cite[SectionII-A]{fong2020opmulti}
An $\left(n, k_{\mathrm{v}}, k_{\mathrm{u}}, T_{\mathrm{v}}, T_{\mathrm{u}}\right)_{\mathbb{F}} $-streaming code consists of the following:
\begin{enumerate}
\item A sequence of messages $\{\mathbf{s}_i\}_{i=0}^{\infty}$ which consists of less urgent and urgent messages, i.e.,  $\mathbf{s}_i=[\mathbf{v}_i,\mathbf{u}_i]$, where  $\mathbf{v}_{i} \in \mathbb{F}^{k_{\mathrm{v}}} $, and $\mathbf{u}_{i} \in \mathbb{F}^{k_{\mathrm{u}}} $.
\item An encoding function  $f_{i}: \underbrace{\mathbb{F}^{k_{\mathrm{u}}+k_{\mathrm{v}}} \times \ldots \times \mathbb{F}^{k_{\mathrm{u}}+k_{\mathrm{v}}}}_{i+1 \text { times }} \rightarrow \mathbb{F}^{n} $ for each  $i \in \mathbb{Z}_{+} $, where  $f_{i}$  is used by the source at time  i  to encode $ \mathbf{u}_{i}$  and  $\mathbf{v}_{i} $ such that
\begin{equation}
\mathbf{x}_{i}=f_{i}\left(\mathbf{s}_0,\mathbf{s}_1,\ldots,\mathbf{s}_i\right) .
\end{equation}
\item A decoding function for less urgent messages
\begin{equation}
\varphi_{i+T_{\mathrm{v}}}^{(\mathrm{v})}: \underbrace{\mathbb{F}^{n} \cup\{*\} \times \ldots \times \mathbb{F}^{n} \cup\{*\}}_{i+T_{\mathrm{v}}+1 \text { times }} \rightarrow \mathbb{F}^{k_{\mathrm{v}}}
\end{equation}
for each  $i \in \mathbb{Z}_{+} $, where  $\varphi_{i+T_{\mathrm{v}}}^{(\mathrm{v})} $ is used by the destination at time  $i+T_{\mathrm{v}}$  to estimate  $\mathbf{v}_{i} $ such that
\begin{equation}
\hat{\mathbf{v}}_{i}=\varphi_{i+T_{\mathrm{v}}}^{(\mathrm{v})}\left(\mathbf{y}_{0}, \mathbf{y}_{1}, \ldots, \mathbf{y}_{i+T_{\mathrm{v}}}\right) .  
\end{equation} 
\item A decoding function for urgent messages
\begin{equation}
\varphi_{i+T_{\mathrm{u}}}^{(\mathrm{u})}: \underbrace{\mathbb{F}^{n} \cup\{*\} \times \ldots \times \mathbb{F}^{n} \cup\{*\}}_{i+T_{\mathrm{u}}+1 \text { times }} \rightarrow \mathbb{F}^{k_{\mathrm{u}}}
\end{equation}
for each  $i \in \mathbb{Z}_{+} $, where  $\varphi_{i+T_{\mathrm{u}}}^{(\mathrm{u})} $ is used by the destination at time  $i+T_{\mathrm{u}} $ to estimate  $\mathbf{u}_{i}$  according to
\begin{equation}
\hat{\mathbf{u}}_{i}=\varphi_{i+T_{\mathrm{u}}}^{(\mathrm{u})}\left(\mathbf{y}_{0}, \mathbf{y}_{1}, \ldots, \mathbf{y}_{i+T_{\mathrm{u}}}\right) .
\end{equation}
\end{enumerate}
\end{Definition}

The formal definition of erasure sequence is given below.

\begin{Definition}
An erasure sequence is a binary sequence denoted by \( e^{\infty} \triangleq\left\{e_{i}\right\}_{i=0}^{\infty} \), where
$$e_{i} = \mathbf{1}\{\text{erasure occurs at time}\, i\}.$$
A  $(W, B, N)$-erasure sequence is an erasure sequence $ e^{\infty} $ that satisfies the following: For each  $i \in \mathbb{Z}_{+}$ and any window
$$\mathcal{W}_{i} \triangleq\{i, i+1, \ldots, i+W-1\},$$
either $ N<\sum_{\ell \in \mathcal{W}_{i}} e_{\ell} \leq B  $ holds with all the 1's in $  \left(e_{i}, e_{i+1}, \ldots, e_{i+W-1}\right) $ occupying consecutive positions or  $\sum_{\ell \in \mathcal{W}_{i}} e_{\ell} \leq N $ holds with no restriction on the positions of 1's. In other words, a  $(W, B, N)$-erasure sequence introduces either one burst erasure with length no longer than  B  or multiple arbitrary erasures with a total count no larger than  N  in any window  $\mathcal{W}_{i}, i \in \mathbb{Z}_{+}$. The set of $(W, B, N)$-erasure sequences is denoted by $\Omega_{(W, B, N)}^{\infty}$ .\\

\end{Definition}

\begin{Definition}
The mapping  $g_{n}: \mathbb{F}^{n} \times\{0,1\} \rightarrow \mathbb{F}^{n} \cup\{*\} $ of the erasure channel is defined as
\begin{equation}
\label{mapping_g_i}
g_{n}(\mathbf{x}, e)=\left\{\begin{array}{ll}
\mathbf{x} & \text { if } e=0, \\
* & \text { if } e=1.
\end{array}\right.
\end{equation}
For any erasure sequence $e^\infty$ and any $\left(n, k_{\mathrm{v}}, k_{\mathrm{u}}, T_{\mathrm{v}}, T_{\mathrm{u}}\right)_{\mathbb{F}}$-streaming code, the following input-output relation holds for the erasure channel for each  $i \in \mathbb{Z}_{+} :$
\begin{equation}
\mathbf{y}_{i}=g_{n}\left(\mathbf{x}_{i}, e_{i}\right) .
\end{equation}
\end{Definition}

Without loss of generality, we assume
\begin{equation}
    W>T_{\mathrm{v}}>T_{\mathrm{u}}>B>N\geq1.
\end{equation}
\begin{Definition}
\label{Df_WBN}
An $\left(n, k_{\mathrm{v}}, k_{\mathrm{u}}, T_{\mathrm{v}}, T_{\mathrm{u}}\right)_{\mathbb{F}}$-streaming code is said to be $(W,B,N)$-achievable if the following holds for any $(W, B, N) \text {-erasure sequence } e^{\infty} \in \Omega_{(W, B, N)}^{\infty}$: \\
For all  $i \in \mathbb{Z}_{+}$ and all  $\mathbf{s}_i=\left[\mathbf{v}_{i}, \mathbf{u}_{i}\right] \in \mathbb{F}^{k_{\mathrm{v}}+k_{\mathrm{u}}}$ , we have
\begin{equation}
\left[\hat{\mathbf{v}}_{i}, \hat{\mathbf{u}}_{i}\right]=\left[\mathbf{v}_{i} ,\mathbf{u}_{i}\right]
\end{equation}
where
$$\hat{\mathbf{v}}_{i}=\varphi_{i+T_{\mathrm{v}}}^{(\mathrm{v})}\left(g_{n}\left(\mathbf{x}_{0}, e_{0}\right), \ldots, g_{n}\left(\mathbf{x}_{i+T_{\mathrm{v}}}, e_{i+T_{\mathrm{v}}}\right)\right)$$
and
$$\hat{\mathbf{u}}_{i}=\varphi_{i+T_{\mathrm{u}}}^{(\mathrm{u})}\left(g_{n}\left(\mathbf{x}_{0}, e_{0}\right), \ldots, g_{n}\left(\mathbf{x}_{i+T_{\mathrm{u}}}, e_{i+T_{\mathrm{u}}}\right)\right).$$
\end{Definition}

\begin{Remark}
    Definition \ref{Df_WBN} serves as a direct extension of the definition of \((W,B,N)\)-achievable streaming codes originally developed for single streams. Further details on this subject can be found in Section \uppercase\expandafter{\romannumeral1}-C of \cite{domanovitz2021anexplicit}.
\end{Remark}

\begin{Definition}
    A region $\mathcal{A}\in \mathbb{R}^{2}_{+}$ is said as achievable region if for any rate pair $(R_{\mathrm{v}},R_{\mathrm{u}}) \in \mathcal{A}$, there exists an $\left(n, k_{\mathrm{v}}, k_{\mathrm{u}}, T_{\mathrm{v}}, T_{\mathrm{u}}\right)_{\mathbb{F}}$-streaming code which is $(W,B,N)$-achievable, such that $\frac{k_{\mathrm{v}}}{n}\geq R_{\mathrm{v}}$, $\frac{k_{\mathrm{u}}}{n}\geq R_{\mathrm{u}}$.
\end{Definition}

Next, we elucidate the definition of $\left(n, k_{\mathrm{v}}, k_{\mathrm{u}}, T_{\mathrm{v}}, T_{\mathrm{u}}\right)_{\mathbb{F}} $-block code.
    
\begin{Definition}
An  $\left(n, k_{\mathrm{v}}, k_{\mathrm{u}}, T_{\mathrm{v}}, T_{\mathrm{u}}\right)_{\mathbb{F}} $-block code consists of the following:
\begin{enumerate}
\item A vector of $(k_{\mathrm{v}}+k_{\mathrm{u}})$ symbols $\vec{s}=[\vec{v},\vec{u}]$ in $\mathbb{F}$, where   $\vec{v}=[v[0],v[1],\ldots,v[k_{\mathrm{v}}-1]]$, and  $\vec{u}=[u[0],u[1],\ldots,u[k_{\mathrm{u}}-1]]$.
\item A generator matrix  $\mathbf{G} \in \mathbb{F}^{\left(k_{\mathrm{v}}+k_{\mathrm{u}}\right) \times n} .$ The source codeword is generated according to
\begin{equation}
[x[0], x[1], \dots, x[n-1]]=[\vec{v}, \vec{u}] \mathbf{G}
\end{equation}

\item A decoding function for less urgent symbols
$$\varphi_{i+T_{\mathrm{v}}}^{(\mathrm{v})}: \underbrace{\mathbb{F} \cup\{*\} \times \ldots \times \mathbb{F} \cup\{*\}}_{\min \left\{i+T_{\mathrm{v}}+1, n\right\} \text { times }} \rightarrow \mathbb{F}$$
for each  $i \in\left\{0,1, \ldots, k_{\mathrm{v}}-1\right\} $, where $ \varphi_{i+T_{\mathrm{v}}}^{(\mathrm{v})} $ is used by the destination at time $\min\{i+T_{\mathrm{v}},n-1\} $  to estimate  $v[i] $ according to
$$\hat{v}[i]=\varphi_{i+T_{\mathrm{v}}}^{(\mathrm{v})}\left(y[0], y[1], \ldots, y\left[\min \left\{i+T_{\mathrm{v}}, n-1\right\}\right]\right) .
$$
\item A decoding function for urgent symbols
$$\varphi_{i+T_{\mathrm{u}}}^{(\mathrm{u})}: \underbrace{\mathbb{F} \cup\{*\} \times \ldots \times \mathbb{F} \cup\{*\}}_{\min \left\{k_{\mathrm{v}}+i+T_{\mathrm{u}}+1, n\right\} \text { times }} \rightarrow \mathbb{F}
$$
for each  $i \in\left\{0,1, \ldots, k_{\mathrm{u}}-1\right\} $, where $ \varphi_{i+T_{\mathrm{u}}}^{(\mathrm{u})} $ is used by the destination at time  $\min\{k_{\mathrm{v}}+i+T_{\mathrm{u}},n-1\} $  to estimate $u[i]$  according to
\small{
$$
\hat{u}[i]=\varphi_{i+T_{\mathrm{u}}}^{(\mathrm{u})}\left(y[0], y[1], \ldots, y\left[\min \left\{k_{\mathrm{v}}+i+T_{\mathrm{u}}, n-1\right\}\right]\right).$$}
\end{enumerate}
\end{Definition}

\begin{Definition}
\label{Df_WBN_block}
An $\left(n, k_{\mathrm{v}}, k_{\mathrm{u}}, T_{\mathrm{v}}, T_{\mathrm{u}}\right)_{\mathbb{F}}$-block code is said to be $(W,B,N)$-achievable if the following holds for any $(W, B, N) \text {-erasure sequence } e^{n} \in \Omega_{(W, B, N)}^{n}$:

Let $y[i]\triangleq g_{1}(x[i], e_i)$ be the symbol received by the destination at time $i$ for each $i\in \{0,1,\dots,n-1\}$, where $g_1$ is as defined in \eqref{mapping_g_i}. Then, we have  $\hat{v}[i]=v[i]$ for all $i\in \{0,1,\dots,k_{\mathrm{v}}-1\}$ and all $v[i]\in \mathbb{F}$, and $\hat{u}[i]=u[i]$ for all $i\in \{0,1,\dots,k_{\mathrm{u}}-1\}$ and all $u[i]\in \mathbb{F}$, where $\hat{v}[i]$ and $\hat{u}[i]$ are as defined in Definition 6.
\end{Definition}

\begin{Lemma}
Given an \(\left(n, k_{\mathrm{v}}, k_{\mathrm{u}}, T_{\mathrm{v}}, T_{\mathrm{u}}\right)_{\mathbb{F}}\)-block code that is \((W, B, N)\)-achievable, there exits a corresponding \(\left(n, k_{\mathrm{v}}, k_{\mathrm{u}}, T_{\mathrm{v}}, T_{\mathrm{u}}\right)_{\mathbb{F}}\)-streaming code that is \((W, B, N)\)-achievable. More specifically, given that \(\mathbf{G}\) is the upper triangular-shaped generator matrix of the \(\left(n, k_{\mathrm{v}}, k_{\mathrm{u}}, T_{\mathrm{v}}, T_{\mathrm{u}}\right)_{\mathbb{F}}\)-block code where \(g_{i,j}\) is the entry situated in row \(i\) and column \(j\) of \(\mathbf{G}\), we can construct an \(\left(n, k_{\mathrm{v}}, k_{\mathrm{u}}, T_{\mathrm{v}}, T_{\mathrm{u}}\right)_{\mathbb{F}}\)-streaming code as follows. For each \(l \in \{0, 1, \dots, n - 1\}\), we construct the following matrices. Here, $k= k_{\mathrm{v}}+k_\mathrm{u}$.
\begin{itemize}
  \item If \(0 \leq l \leq n - k\),

\scalebox{0.96}{$
\mathbf{G}_l \triangleq \left[ 
  \mathbf{0}^{k \times l} \text{diag}(g_{0,l}, g_{1,l+1}, \dots, g_{k-1,l+k-1}) \mathbf{0}^{k \times (n - k - l)}
\right].$}

 \item  If \(n - k \leq l \leq n - 1\),
\begin{equation*}
\mathbf{G}_l \triangleq \left[\begin{array}{cc}
    \multirow{2}{*}{$\mathbf{0}^{k \times l}$} & \text{diag}(g_{0,l}, g_{1,l+1}, \dots, g_{n-1,l+n-1}) \\
    & \mathbf{0}^{(k - n + l) \times (n - l)}
\end{array}\right].
\end{equation*}
\end{itemize}
We note that \(\mathbf{G}= \sum_{l=0}^{n-1} \mathbf{G}_l\). In particular, if we let \(\mathbf{s}_i \triangleq [s_i[0],s_i[1] ,\dots, s_i[k - 1]]=[v_i[0],\dots,v_i\left[k_\mathrm{v}-1\right],u_i[0],\dots,u_i[k_\mathrm{u}-1]]\) and let
\begin{equation*} 
\begin{aligned}
   \relax[x_i[0], x_{i+1}[1], \dots, x_{i+n-1}[n - 1]] \triangleq\\
   [s_i[0], s_{i+1}[1], \cdots, s_{i+k-1}[k - 1]] \mathbf{G}, 
\end{aligned}
\end{equation*}
i.e., we apply diagonal interleaving for all \(i \in \mathbb{Z}_+ \cup \{0\}\), then the symbols generated at time \(i\) by the \(\left(n, k_{\mathrm{v}}, k_{\mathrm{u}}, T_{\mathrm{v}}, T_{\mathrm{u}}\right)_{\mathbb{F}}\)-streaming code are
\begin{equation}
   \mathbf{x}_i \triangleq [x_i[0], x_i[1], \dots , x_i[n - 1]]=\sum_{l=0}^{n-1}\mathbf{s}_{i-l}\mathbf{G}_l,
\end{equation}
where $\mathbf{s}_{-1} = \mathbf{s}_{-2} = \dots = \mathbf{s}_{-n+1} = \mathbf{0}^{1\times k}$ by convention.
\end{Lemma}
\begin{proof}
    The detailed proof is in Appendix A.
\end{proof}
This lemma serves as a straight extension of the existing result for single flows \cite[Lemma 1]{fong2019optimal}. It allows us to focus our attention solely on \(\left(n, k_{\mathrm{v}}, k_{\mathrm{u}}, T_{\mathrm{v}}, T_{\mathrm{u}}\right)_{\mathbb{F}}\)-block codes, obviating the need for separate analyses of streaming codes.

\subsection{Revisit of the Explicit Construction of $(W,B,N)$-Achievable Rate-Optimal Block Code in \cite[Section \uppercase\expandafter{\romannumeral2}-B]{domanovitz2021anexplicit}}

\begin{figure}[!b]
\centerline{\includegraphics[width=1\linewidth]{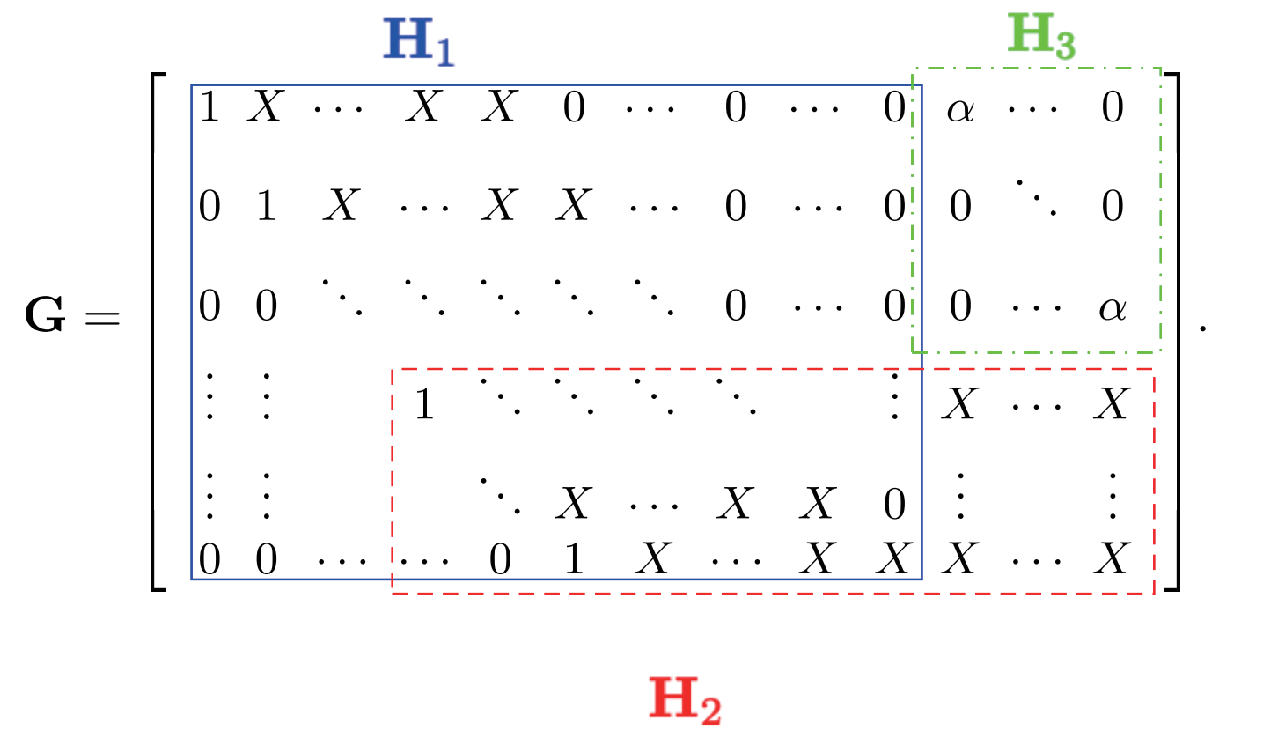}}
\caption{Generator matrix of the block code in \cite{domanovitz2021anexplicit}}\label{Fig_geneM}
\end{figure}

In this subsection, we revisit the explicit construction of $(W,B,N)$-achievable rate-optimal block code in \cite[Section \uppercase\expandafter{\romannumeral2}-B]{domanovitz2021anexplicit}, which is a key building block of our new code.
The generator matrix \( \mathbf{G} \) for the \( (n, k, T)_{\mathbb{F}}\)-block code, as described in \cite[Section \uppercase\expandafter{\romannumeral2}-B]{domanovitz2021anexplicit}, is illustrated in Fig. \ref{Fig_geneM}. Here, \( X \) and \( \alpha \) represent elements from \( \mathbb{F}_{q} \) and \( \mathbb{F}_{q^{2}} \backslash \mathbb{F}_{q} \), respectively, with \( q = O(T) \). Furthermore, the relationship between \( k \), \( n \), and \( T \) is specified as follows:
\begin{equation}
\begin{aligned}
k &= T - N + 1, \\
n &= k + B.
\end{aligned}
\end{equation}

This ensures that the code rate \( \frac{k}{n} \) satisfies the capacity constraint described in Eq. \eqref{Eqn_capa_MS}.

The codeword can be expressed as:
\begin{equation}
    [x[0], x[1], \ldots, x[T-N+B]] = [s[0], s[1], \ldots, s[T-N]] \mathbf{G}.
\end{equation}

The generator matrix $\mathbf{G}$ is composed of the following three blocks. Fig. \ref{Fig_geneM} presents the explicit form of $\mathbf{G}$.
\begin{itemize} 
\item $\mathbf{H}_{1}$:  The upper left $k \times(k+N-1)$  sub-matrix of  $\mathbf{G}$ .
\item $\mathbf{H}_{2}$:  The lower right  $(k-(B-N+1)) \times   (n-(B-N+1))$  sub-matrix of  $\mathbf{G}$ .
\item $\mathbf{H}_{3}$: The upper right  $(B-N+1) \times(B-N+1)$  sub-matrix of  $\mathbf{G}$ .
\end{itemize}
We revisit certain properties of this construction as detailed in \cite[Section \uppercase\expandafter{\romannumeral2}-B]{domanovitz2021anexplicit}.

\begin{Lemma}
The block matrices $\mathbf{H}_{1}$ and $\mathbf{H}_{2}$ have the following properties:
\begin{enumerate}
\item $\mathbf{H}_{1}$ in Fig. \ref{Fig_geneM} is a generator matrix of a $(k+N-1, k)$ MDS code over $\mathbb{F}_{q}$.
\item $\mathbf{H}_{2}$ in Fig. \ref{Fig_geneM} is a generator matrix of an $(n-(B-N+1), k-(B-N+1))$ MDS code over $\mathbb{F}_{q}$.
\end{enumerate}
\end{Lemma}

\begin{Lemma}
\label{ori_MDS_proposed_lemma}
Any set of $k$ consecutive columns in the matrix $\mathbf{G}$ is linearly independent.
\end{Lemma}
\begin{proof}
 The detailed proof is in Appendix B. 
\end{proof}

\begin{figure}[!b]
    \centering
    \begin{minipage}[t]{\columnwidth}
        \centering
        \includegraphics[width=0.9\textwidth]{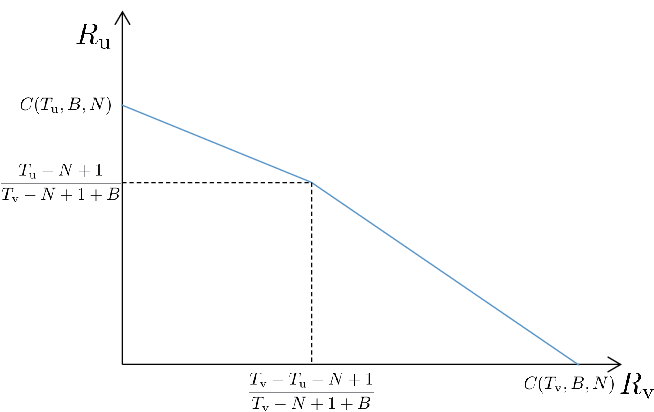} 
        \caption*{(a) Case $B\geq 2N-1$}
        \label{fig:image1}
    \end{minipage}\\[1em] 

    \begin{minipage}[t]{\columnwidth}
        \centering
        \includegraphics[width=0.9\textwidth]{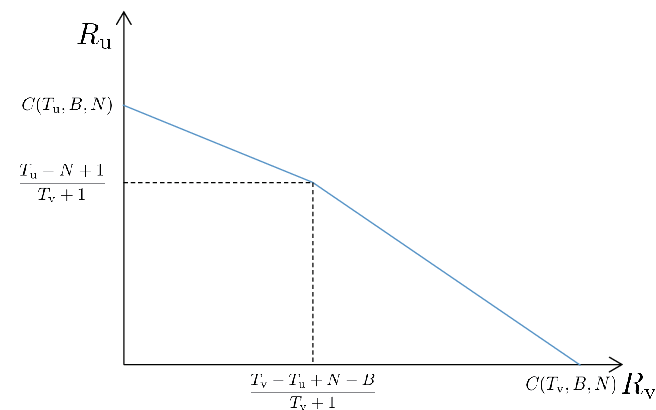} 
        \caption*{(b) Case $B< 2N-1$}
        \label{fig:image2}
    \end{minipage}
    \caption{Achievable region $\mathcal{R}$ when $T_{\mathrm{v}}>T_{\mathrm{u}}+B$.}
    \label{achievabl_region_R_combined}
\end{figure}

\section{Main results}
The main results of this work consist of two parts. In the first part, we propose a non-trivial achievable region $\mathcal{R}$ for multiplexed streaming codes in sliding window model. Fig. 2 presents the visual area of $\mathcal{R}$, and the specific mathematical form of $\mathcal{R}$ is illustrated in Theorem 1.

\begin{Theorem}
\label{thm_R_boundary}
    When $T_{\mathrm{v}}>T_{\mathrm{u}}+B$, define
\begin{equation}
\resizebox{0.99\columnwidth}{!}{%
$
\mathcal{R}\triangleq \left\{(R_{\mathrm{v}},R_{\mathrm{u}}) \,\middle|\,  
\begin{aligned}
&(C(T_{\mathrm{u}},B,N)-b)R_{\mathrm{v}}+aR_{\mathrm{u}}-aC(T_{\mathrm{u}},B,N)\leq 0, \\
&bR_{\mathrm{v}}+(C(T_{\mathrm{v}},B,N)-a)R_{\mathrm{u}}-(C(T_{\mathrm{v}}-a))b\leq 0,\\
&(R_{\mathrm{v}}, R_{\mathrm{u}}) \in \mathbb{R}^2_+.
\end{aligned}
\right\}$
}
\end{equation}
Here, 
\begin{equation}
\resizebox{.99\hsize}{!}{
    \ensuremath{
        (a,b) = \left\{
        \begin{aligned}
            &\left(\frac{T_{\mathrm{v}}-T_{\mathrm{u}}-N+1}{T_{\mathrm{v}}-2N+2+B}, \frac{T_{\mathrm{u}}-N+1}{T_{\mathrm{v}}-2N+2+B}\right), && \text{if } B \geq 2N-1, \\
            &\left(\frac{T_{\mathrm{v}}-T_{\mathrm{u}}+N-B}{T_{\mathrm{v}}+1}, \frac{T_{\mathrm{u}}-N+1}{T_{\mathrm{v}}+1}\right), && \text{if } B \leq 2N-1.
        \end{aligned}
        \right.
    }
}
\end{equation}
Then, region $\mathcal{R}$ is an achievable region covering the trivial achievable region $\mathcal{B}$, where $\mathcal{B}$ denotes the trivial achievable region enclosed by the line though the points $(C(T_{\mathrm{v}},B,N),0)$ and $(0,C(T_{\mathrm{u}},B,N))$ with the positive semiaxis. 
\end{Theorem}
\begin{proof}
    The proof is in Appendix C.
\end{proof}
\begin{Remark}
    The achievability of region $\mathcal{R}$ is derived by the convex combination of the non-trivial corner point $(a,b)$ stated in Theorem 1 with other two extreme points. In section IV, we propose a novel merging method and construct explicit constructions of $(W, B, N)$-achievable block codes, whose rate pair corresponds to $(a,b)$.
\end{Remark}


In the second part, we investigate the property of region $\mathcal{R}$, specifically, the following conclusion presented in Remark 3.



\begin{Remark}
    We prove that all the rate pairs achievable through the merging method are strictly contained in $\mathcal{R}$, which indicates that $\mathcal{R}$ is the largest achievable region derived by merging method. For further details, please refer to Theorem  \ref{omega_in_R} in Section VI.
\end{Remark}

Moreover, we present a comparison with existing schemes in Section IV-D. It indicates that $\mathcal{R}$ includes all the rate pairs corresponding to the existing benchmarks. When sliding window model reduced to the burst channel, i.e., $B>N=1$, $\mathcal{R}$ is equivalent to the capacity region in \cite{fong2020opmulti}.

\begin{figure*}[htbp]
\centerline{\includegraphics[width=1\linewidth]{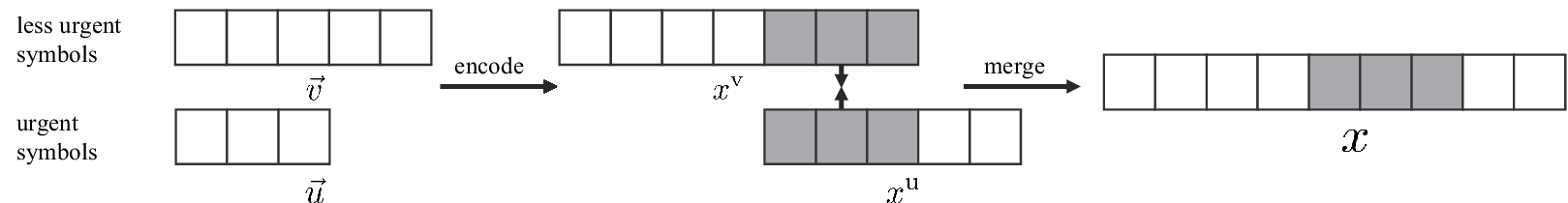}}
\caption{The flow chart of the encoding steps based on merging method. The shadow blocks indicate the messages that overlap with others.}
\label{Fig_flowchart}
\end{figure*}

\section{Construction}
We first introduce a novel merging method for encoding the two streams of messages, leveraging the \((W, B, N)\)-achievable rate-optimal block code as the foundation.

Pursuant to this strategic method, we offer explicit constructions of $(W,B,N)$-achievable block codes, which proves the achievability of the non-trivial corner point $(a,b)$ in aforementioned section. The detailed proof is presented in Section V.

\subsection{Merging Method}
Let \(\vec{v}\) and \(\vec{u}\) represent the less urgent and urgent messages of lengths \(k_{\mathrm{v}}\) and \(k_{\mathrm{u}}\), respectively. Fig. \ref{Fig_flowchart} presents the flowchart of the overall merging method. Our encoding strategy unfolds in the following manner:
\begin{itemize}
    \item[] \textbf{Step 1:} Utilize the construction of \((W,B,N)\)-achievable rate-optimal block codes as detailed in \cite[Section \uppercase\expandafter{\romannumeral2}-B]{domanovitz2021anexplicit} to encode both \(\vec{v}\) and \(\vec{u}\) separately. The encoded messages can be represented as:
    \begin{equation}
    \label{eq_lessurg}
    \vec{v}G_1 = \left[x^{(\text{v})}[0], x^{(\text{v})}[1], \ldots, x^{(\text{v})}[k_{\mathrm{v}}+B-1]\right],
    \end{equation}
    \begin{equation}
    \label{eq_urgent}
    \vec{u}G_2 = \left[x^{(\text{u})}[0], x^{(\text{u})}[1], \ldots, x^{(\text{u})}[k_{\mathrm{u}}+B-1]\right],
    \end{equation}
    where \(G_1\) and \(G_2\) are generator matrices corresponding to the construction from \cite[Section \uppercase\expandafter{\romannumeral2}-B]{domanovitz2021anexplicit}. The parameters \(k_{\mathrm{u}}\) and \(k_{\mathrm{v}}\) are determined by:
    \begin{equation}
    \begin{aligned}
    k_{\mathrm{u}} &= T_{\mathrm{u}}' - N + 1, \\
    k_{\mathrm{v}} &= T_{\mathrm{v}}' - N + 1,
    \end{aligned}
    \end{equation}
    where \(T_{\mathrm{u}}' \leq T_{\mathrm{u}}\) and \(T_{\mathrm{v}}' \leq T_{\mathrm{v}}\).

    \item[] \textbf{Step 2:} Merge the separately encoded messages to following form with a certain size $m$:
    \begin{equation}
    \begin{aligned}
        &[x[0],x[1],\dots,x[n-1]] \\
        = &[x^{(\mathrm{v})}[0],\ldots, x^{(\mathrm{v})}[k_{\mathrm{v}}+B-m-1], 
        x^{(\mathrm{v})}[k_{\mathrm{v}}+B-m] \\
        +& x^{(\mathrm{u})}[0],\ldots, x^{(\mathrm{v})}[k_{\mathrm{v}}+B-1]+x^{(\mathrm{u})}[m-1], 
        x^{(\mathrm{u})}[m], \\
        &\ldots, 
        x^{(\mathrm{u})}[k_{\mathrm{u}}+B-1]],
    \end{aligned}
\end{equation}
where $m$ denotes the size of overlapped symbols between the two sources of messages. 

    That is
    \begin{equation}
    \begin{split}
    &[x[0],x[1],\dots,x[n-1]]\\
    =&[v[0],v[1],\dots,v[k_{\mathrm{v}}-1],u[0],u[1],u[k_{\mathrm{u}}-1]]\mathbf{G},
    \end{split}
    \end{equation}
    where 
    \begin{equation}
    \label{G_merge}
    \mathbf{G}=\begin{array}{c@{\hspace{-5pt}}l}
    \left[\begin{array}{ccc}
    G_{1[1:i]}&G_{1 [i+1:k_{\mathrm{v}}+B]}&\mathbf{0}\\
    \mathbf{0}&G_{2[1:m]}&G_{2[m+1:k_{\mathrm{u}}+B]}\
     \end{array}\right],
      \end{array}
      \end{equation}
    $i=k_{\mathrm{v}}+B-m.$
\end{itemize}
\begin{Remark} The encoding in Step 1 adopts a \((W, B, N)\)-achievable rate-optimal block code for both urgent and less urgent messages. New deadlines \(T_{\mathrm{u}}'\) and \(T_{\mathrm{v}}'\) are set for the urgent and less urgent messages, respectively, under the constraint that \(T_{\mathrm{u}}' \leq T_{\mathrm{u}}\) and \(T_{\mathrm{v}}' \leq T_{\mathrm{v}}\). If either \(T_{\mathrm{u}}' > T_{\mathrm{u}}\) or \(T_{\mathrm{v}}' > T_{\mathrm{v}}\), the encoded messages can not be decoded within the specified deadlines.        
    \end{Remark}
\subsection{Explicit Constructions Based on Merging Method}

\subsubsection{The construction when $B\geq 2N-1$}
$ $

When $T_{\mathrm{v}}>T_{\mathrm{u}}+B>T_{\mathrm{u}}>B$, we propose an $(n,k_{\mathrm{v}},k_{\mathrm{u}},T_{\mathrm{v}},T_{\mathrm{u}})_{\mathbb{F}}$ $(W,B,N)$-achievable block code based on merging method. The explicit construction is as follows:

Taking the merging size $m=B$, we define:
\begin{equation}
    \begin{gathered}
         k_{\mathrm{u}}=T_{\mathrm{u}}'-N+1,\\
       k_{\mathrm{v}}=T_{\mathrm{v}}-T_{\mathrm{u}}'-N+1,\\
         n=k_{\mathrm{u}}+k_{\mathrm{v}}+2B-m=T_{\mathrm{v}}-2N+2+B.
    \end{gathered}
\end{equation}
Consequently, the obtained sum rate is given by:
\begin{equation}
R_{\mathrm{u}} + R_{\mathrm{v}} = \frac{T_{\mathrm{v}} - 2N + 2}{T_{\mathrm{v}} - 2N + 2 + B}.
\end{equation}

\begin{figure}[!b]
\centerline{\includegraphics[width=1\linewidth]{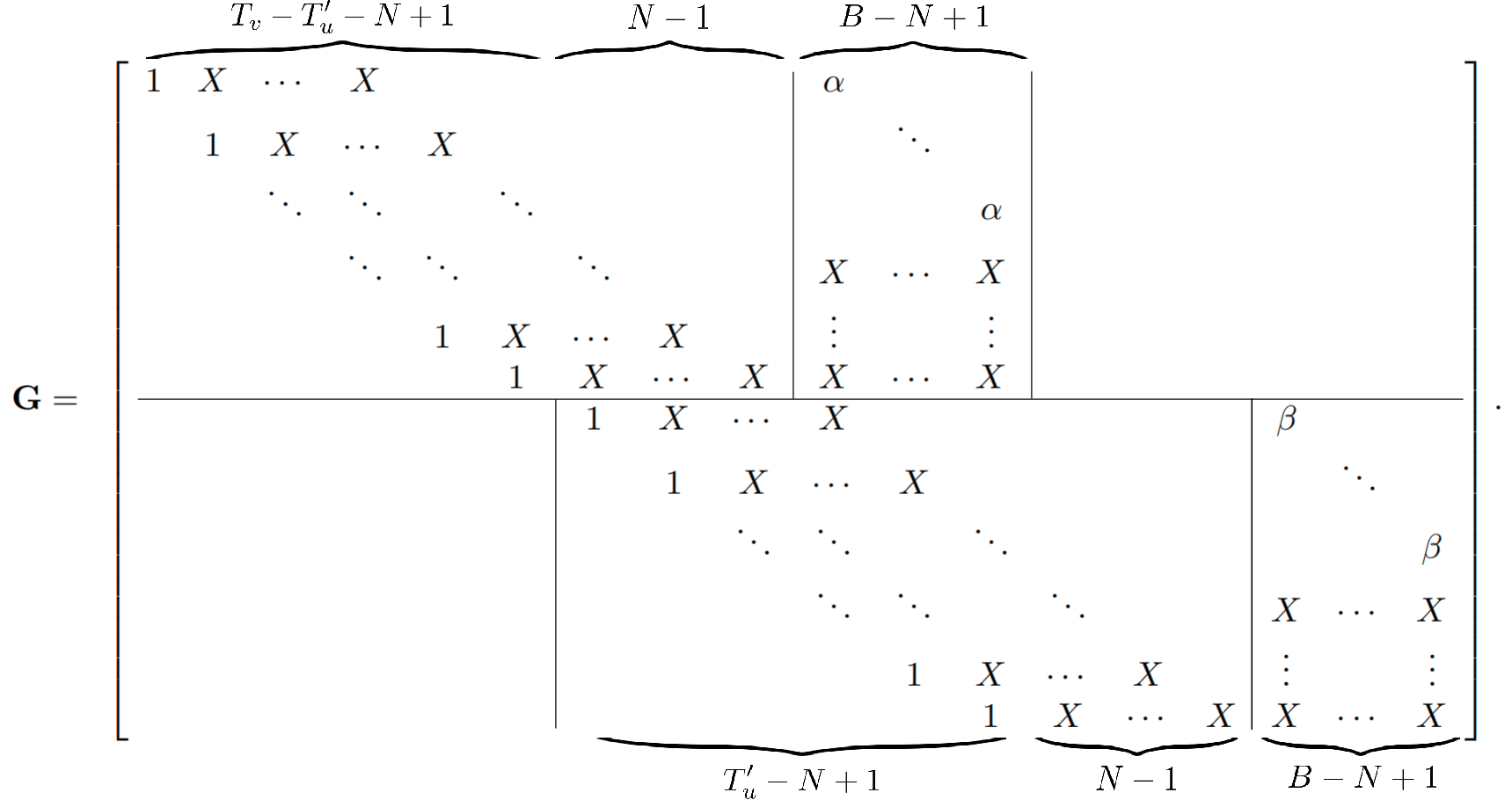}}
\caption{The generator matrix $\mathbf{G}$ of the proposed code when $B\geq 2N-1$}
\label{generator_matrix_1}
\end{figure}

The codeword is as follows:
\begin{equation}
\label{cons_sizeB}
\begin{split}
[x[0],x[1],\dots,x[T_{\mathrm{v}}-2N+1+B]]=\\
[v[0],v[1],\dots,v[T_{\mathrm{v}}-T_{\mathrm{u}}'-N],u[0],u[1],u[T_{\mathrm{u}}'-N]]\mathbf{G},
\end{split}
\end{equation}
\begin{equation}
\left[x^{(\mathrm{v})}[0], x^{(\mathrm{v})}[1], \ldots, x^{(\mathrm{v})}\left[T_{\mathrm{v}}-T_{\mathrm{u}}'-N+B\right]\right]=\vec{v} G_1,
\end{equation}
\begin{equation}
\left[x^{(\mathrm{u})}[0], x^{(\mathrm{u})}[1] ,\ldots, x^{(\mathrm{u})}\left[T_{\mathrm{u}}'-N+B\right]\right]=\vec{u}G_2,
\end{equation}
where 
\begin{equation}
\mathbf{G}=\begin{array}{c@{\hspace{-5pt}}l}
\left[\begin{array}{ccc}
G_{1[1:k_{\mathrm{v}}]}&G_{1 [k_{\mathrm{v}}+1:k_{\mathrm{v}}+B]}&\mathbf{0}\\
\mathbf{0}&G_{2[1:B]}&G_{2[B+1:k_{\mathrm{u}}+B]}\
 \end{array}\right].
  \end{array}
  \end{equation}
Here, $G_1$ and $G_2$, with dimensions $k_{\mathrm{v}} \times (k_{\mathrm{v}} + B)$ and $k_{\mathrm{u}} \times (k_{\mathrm{u}} + B)$ respectively, are generator matrices of $(W, B, N)$-achievable rate-optimal block codes in Fig. \ref{Fig_geneM}. Additionally, Fig. \ref{generator_matrix_1} illustrates the explicit form of generator matrix $\mathbf{G}$.

\begin{Remark}
The element $\alpha$ in the generator matrix $\mathbf{G}$ does not need such an expansive field size as designed in \cite{domanovitz2021anexplicit}. Here, we designate the element $\alpha \in \mathbb{F}_{q}$ within $G_1$, while the element $\beta \in \mathbb{F}_{q^{2}} \backslash \mathbb{F}_{q}$ persists within $G_2$. A more comprehensive discussion will be provided in Section V.
\end{Remark}

\begin{Lemma}
\label{MDS_lemma}
The sub-matrix $\widetilde{G}$, which is generated by the first $T_{\mathrm{v}}-N+1$ columns of $\mathbf{G}$, serves as a generator matrix for a $(T_{\mathrm{v}} - N + 1, T_{\mathrm{v}} - 2N + 2)_{\mathbb{F}_q}$ MDS code.
\end{Lemma}
\begin{proof}
    The detailed proof is in Appendix D.
\end{proof}
Next, we will present a theorem validating that this construction of the $(n,k_{\mathrm{v}},k_{\mathrm{u}},T_{\mathrm{v}},T_{\mathrm{u}})_{\mathbb{F}}$-block code is $(W,B,N)$-achievable.

\begin{Theorem}
\label{thm_Bsize}
   For the case where $B \geq 2N - 1$,  the construction of the $(n,k_{\mathrm{v}},k_{\mathrm{u}},T_{\mathrm{v}},T_{\mathrm{u}})_{\mathbb{F}}$-block code given in \eqref{cons_sizeB}  is $(W,B,N)$-achievable.
\end{Theorem} 
\begin{proof}
The proof is in the next section.
\end{proof}

\subsubsection{The construction when $B<2N-1$}
$ $

\input{construction_N}

\subsection{An Example}

Defining $B = 4$, $N = 2$, $T_{\mathrm{v}} = 12$, and $T_{\mathrm{u}} = 6$, we select $k_{\mathrm{v}} = T_{\mathrm{v}} - T_{\mathrm{u}} - N + 1 = 5$, $k_{\mathrm{u}} = T_{\mathrm{u}} - N + 1 = 5$, and $n = T_{\mathrm{v}} - 2N + 2 + B = 14$. The sum rate in this construction is given by: 
$$\frac{k_{\mathrm{u}}+k_{\mathrm{v}}}{n}=0.7143.$$

Given that $k_{\mathrm{v}} = 5$ and $k_{\mathrm{u}} = 5$, the proportion of urgent messages in the data stream corresponds to a 1:1 ratio with the proportion of less urgent messages. If we encode these two streams separately while maintaining the same ratio, the resulting maximal sum rate is:
$$\frac{2}{\frac{1}{C(T_{\mathrm{v}},B,N)}+\frac{1}{C(T_{\mathrm{u}},B,N)}}=0.632.$$
Clearly, our proposed codes have a relevant gain in sum rate by 13.0\% over separate encoding.

\begin{figure}[!b]
\centerline{\includegraphics[width=1\linewidth]{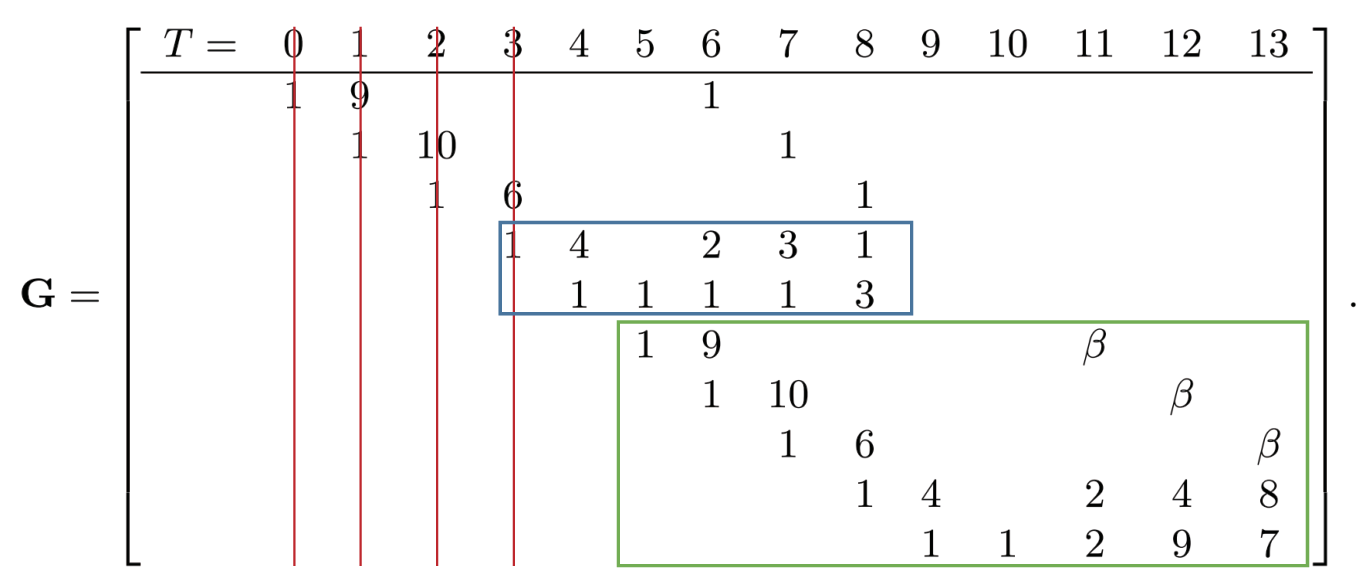}}
\caption{Example: The generator matrix of $(T_{\mathrm{v}},T_{\mathrm{u}},B,N)=(12,6,4,2)$ in channel with burst}
\label{example1}
\end{figure}

\begin{figure}[!t]
\centerline{\includegraphics[width=1\linewidth]{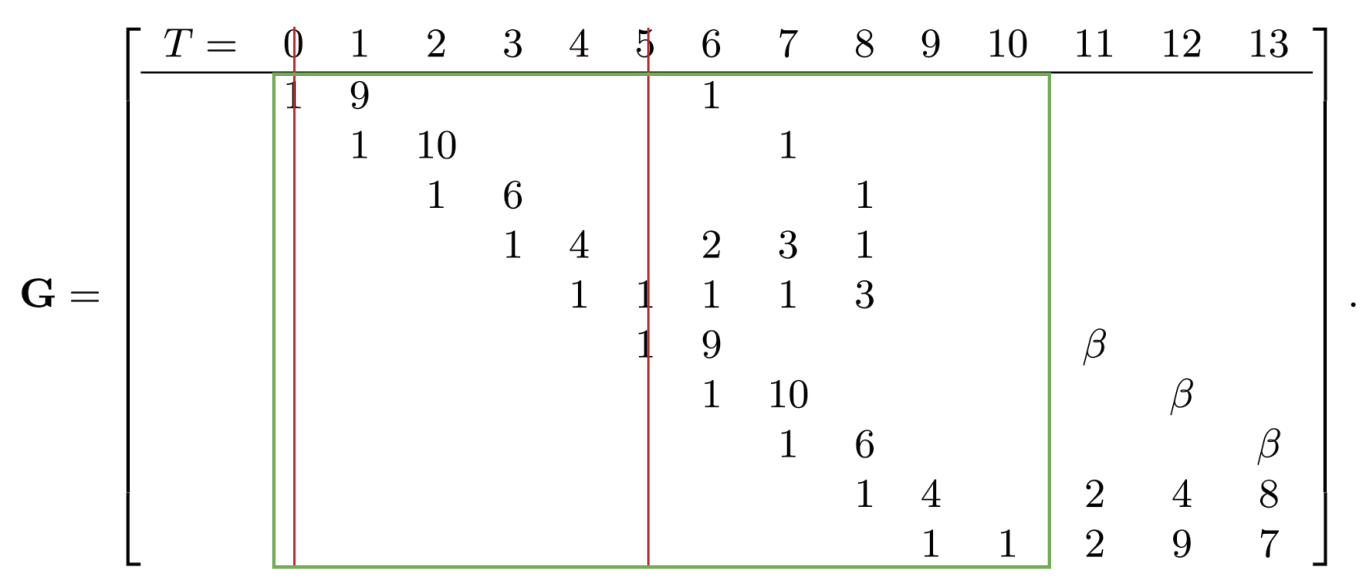}}
\caption{Example: The generator matrix of $(T_{\mathrm{v}},T_{\mathrm{u}},B,N)=(12,6,4,2)$ block code in channel with random erasures}
\label{example2}
\end{figure}

Fig. \ref{example1}, \ref{example2} present the same generator matrix in burst or random erasure channel. We set  $\beta = 0 \cdot 11+1\cdot 11 \in GF(121)\backslash GF(11)$ as an example, the other elements are all in $GF(11)$. Subsequently, we outline the decoding steps in the following two scenarios with burst or random erasures.

\noindent\textbf{-Length-B burst-} 

Suppose a length-4 burst beginning at $i=0$. At this point, $x^{(\mathrm{v})}[4]$ is already known. The matrix enclosed within the green box in Fig. \ref{example1} is a generator matrix of a $(7, 4, 2)$-achievable block code, which enables that urgent symbols can be decoded in time. Consequently, at $i = 11$, we can decode $u[0]$; similarly, at $i = 12$, we decode $u[1]$. This leads to the acquisition of $x^{(\mathrm{u})}[0]$ and $x^{(\mathrm{u})}[1]$ at $i = 12$. By canceling these symbols from $x[5]$ and $x[6]$, we are able to retrieve $x^{(\mathrm{v})}[5]$ and $x^{(\mathrm{v})}[6]$.

Next, consider the matrix within the blue box in Fig. \ref{example1}, which represents a $(6, 2)_{GF(11)}$ MDS code. We can decode $v[4]$ and $v[5]$ using this MDS code, based on the information from $x^{(\mathrm{v})}[4]$ and $x^{(\mathrm{v})}[5]$. As a result of canceling $v[4]$ and $v[5]$ from $x^{(\mathrm{v})}[6]$, we can decode $v[0]$. The cumulative delay for recovering $v[0]$ is 12. Additionally, all the symbols can be decoded in the same discussion within deadlines.

\noindent\textbf{-N random erasures-}

Assume that random erasures occur at $i = 0$ and $i = 5$. By virtue of Lemma \ref{MDS_lemma}, after erasing $x[0]$ and $x[5]$, the initial 9 columns of the matrix within the green box remain linearly independent. Consequently, every information symbol can be represented as a combination of $u[0]$ and a constant, which can be
expressed as follows:
\begin{align}
\label{23}
u[j] = a_j + b_ju[0], \\
\label{24}
v[k] = \widetilde{a}_k + \widetilde{b}_ku[0],
\end{align}
where $a_j, b_j, \widetilde{a}_k, \widetilde{b}_k \in GF(11)$, with $j \in \{1, 2, \ldots, 4\}$ and $k \in \{0, 1, \ldots, 4\}$.

Consider the $12$-th column of the generator matrix $\mathbf{G}$, which leads to the following equation:
\begin{equation}
\label{25}
x[11] = \beta u[0] + 2u[3] + 2u[4].
\end{equation}

Incorporating equations \eqref{23} and \eqref{24} into equation \eqref{25}, given that $\beta \in GF(121)\backslash GF(11)$, the coefficient of $u[0]$ in equation \eqref{25} is non-zero. Hence, we can decode $u[0]$ when $i = 11$. Furthermore, all information symbols can also be successfully decoded.

\subsection{Comparison}
In our work, we examine a scenario where two streams are multiplexed in the sliding window model. The capacity region in this scenario is obviously bounded by two trivial bounds: the rate of the urgent symbols, denoted by $R_{\mathrm{u}} \leq C(T_{\mathrm{u}}, B, N)$, and the sum rate, denoted by $R_{\mathrm{u}} + R_{\mathrm{v}} \leq C(T_{\mathrm{v}}, B, N)$, within which the proposed achievable region $\mathcal{R}$ naturally resides. Given the core focus of this paper, we will next focus on comparing the size of the proposed achievable region $\mathcal{R}$ with the rate pair region derived by existing benchmark schemes. Since there is limited literature on constructions for multiplexed encoding in the sliding window model, there are encoding schemes for specific cases that serve as useful benchmarks.

\begin{figure*}[htbp]
    \centering
    \begin{minipage}{.3\textwidth}
        \centering
        \includegraphics[width=\linewidth]{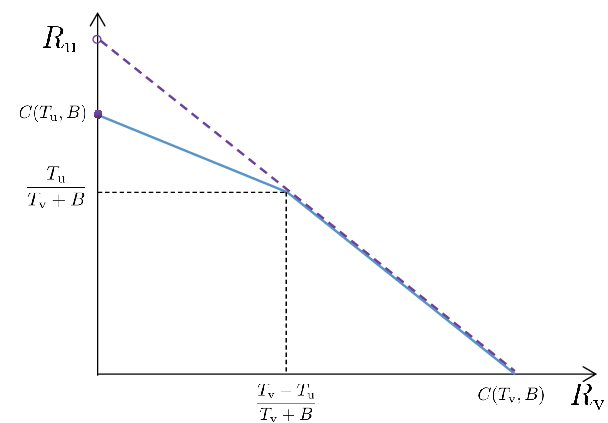}
        \caption*{(a) \(B > N = 1\)}
    \end{minipage}%
    \hfill
    \begin{minipage}{.3\textwidth}
        \centering
        \includegraphics[width=\linewidth]{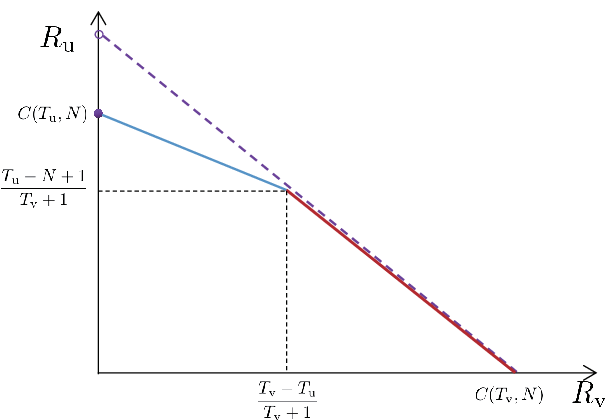}
        \caption*{(b) \(B = N\)}
    \end{minipage}%
    \hfill
    \begin{minipage}{.3\textwidth}
        \centering
        \includegraphics[width=\linewidth]{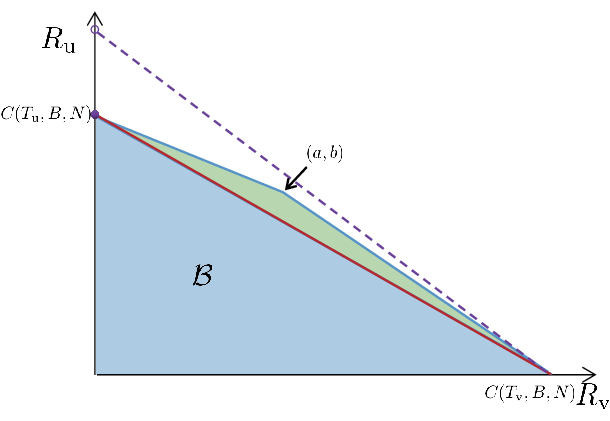}
        \caption*{(c) \(B > N > 1\)}
    \end{minipage}
    \caption{Comparison in the sliding window model under specific conditions. The achievable region $\mathcal{R}$ we proposed is represented by the area enclosed by the blue lines and the coordinate axes.}
    \label{fig:comparison}
\end{figure*}

The benchmarks include:
\begin{itemize}
    \item \textbf{\cite{fong2020opmulti}'s proposed codes in terms of multiplexed erasure codes for burst channel with different decoding delays:} 
    
    \cite{fong2020opmulti} addresses the situation where multiplexed streaming codes are encoded in a burst channel, particularly when \(T_{\mathrm{v}} > T_{\mathrm{u}} + B\). It characterizes the capacity region for this scenario and provides a construction whose rate pair is located at the boundary of the capacity region. This construction is applicable as a benchmark when \(B > N = 1\).

    \item  \textbf{\cite{fong2020threenode}, \cite{Facenda2023multinodes}'s proposed codes in terms of $N$-achievable point-to-point block code with delay spectrum:} 
    
    \cite{fong2020threenode}, \cite{Facenda2023multinodes} introduce an \(N\)-achievable point-to-point \((n,k,\Delta)_\mathbb{F}\)-block code with a delay spectrum : \(\Delta=(T,T-1,\dots,N)\), where \(T \geq N\), \(k \triangleq T - N + 1\), and \(n \triangleq k + N\). The \(i\)-th element in the delay spectrum represents the maximum delay required to decode the \(i\)-th symbol. Therefore, this code can be used as a benchmark for cases where \(B = N\). Specifically, symbols are denoted as \([\vec{v},\vec{u}]=[\underbrace{s[0],\dots,s[i-1]}_{\text{less urgent symbols}}, \underbrace{s[i],\dots,s[k-1]}_{\text{urgent symbols}}]\), with less urgent symbols constrained by a delay of \(T\), and urgent symbols by a delay of \(T - i\) for \(i \leq k - 1\). 

    \item \textbf{\cite{domanovitz2021anexplicit}'s proposed codes in terms of rate-optimal $(W,B,N)$-achievable block codes for the sliding window model in single flow:} 
    
    \cite{domanovitz2021anexplicit} presents an explicit rate-optimal streaming code for a single flow in the sliding window model. For two streams with different delay constraints, each employing the rate-optimal coding construction and then combining them into one stream with varying ratios, this approach can be used as a benchmark for cases where \(B > N > 1\). We refer to this encoding approach as separate encoding.
\end{itemize}
We then compare the aforementioned benchmarks with the achievable region $\mathcal{R}$ that we proposed in the scenario where $T_{\mathrm{v}} > T_{\mathrm{u}} + B > T_{\mathrm{u}}$.
A graphical comparison is shown in Fig. \ref{fig:comparison}, which illustrates the region of the rate pairs derived from the proposed scheme and the benchmarks in specific cases.
Additionally, the two trivial bounds, namely the rate of the urgent symbols, $C(T_{\mathrm{u}}, B, N)$, and the sum rate, $C(T_{\mathrm{v}}, B, N)$, are depicted in Fig. \ref{fig:comparison} as a purple circle and a purple dotted line, respectively.
A more comprehensive analysis of the comparison is provided as follows.

\begin{itemize}
    \item \textbf{\(B > N = 1:\)}
    
    In this scenario, the corner point  \((\frac{T_{\mathrm{v}} - T_{\mathrm{u}}}{T_{\mathrm{v}} + B}, \frac{T_{\mathrm{u}}}{T_{\mathrm{v}} + B})\) in region \(\mathcal{R}\) matches the non-trivial corner point mentioned in \cite{fong2020opmulti}. Furthermore, it is evident that the achievable region $\mathcal{R}$, as shown in Fig. \ref{fig:comparison} (a), is equivalent to the capacity region proposed in \cite{fong2020opmulti} for this scenario. This equivalence demonstrates that our proposed scheme accurately captures the capacity bounds in this special case.

    \item \textbf{\(B = N:\)}
    
    In this scenario, the rate pairs of the point-to-point block code proposed in \cite{fong2020threenode} can be mapped within the region, as indicated by the red line in Fig. \ref{fig:comparison} (b). It is apparent that this line falls within the achievable region \(\mathcal{R}\). Moreover, the sum rate of the non-trivial rate pair $(\frac{T_{\mathrm{v}}-T_{\mathrm{u}}}{T_{\mathrm{v}}+1}, \frac{T_{\mathrm{u}}-N+1}{T_{\mathrm{v}}+1})$ is equivalent to $C(T_{\mathrm{v}},N)$, and one of the boundaries of $\mathcal{R}$, as illustrated in Theorem \ref{thm_R_boundary}, coincides with the trivial sum rate upper bound $C(T_{\mathrm{v}},N)$.
   
    \item \textbf{\(B > N >1:\)}
    
    In this scenario, we compare our approach with separate encoding as shown in Fig. \ref{fig:comparison} (c). The rate pairs obtained through separate encoding are depicted by the red line, which is entirely encompassed within the achievable region \(\mathcal{R}\). The blue region \(\mathcal{B}\) represents the trivial achievable region derived from separate encoding. The size of region \(\mathcal{R}\) is strictly larger than that of \(\mathcal{B}\), with the extended area highlighted in green in Fig. \ref{fig:comparison} (c). Moreover, we can observe there is a gap between the achievable region $\mathcal{R}$ with the trivial upper bounds. It remains an open question to investigate the capacity region \cite{fong2020opmulti}.
\end{itemize}

\subsection{Further Discussion on Sum Rate}
In this section, we delve into a more detailed discussion about sum rate when comparing with separate encoding. We assert that, when $T_{\mathrm{v}}>T_{\mathrm{u}}+B>T_{\mathrm{u}}>B$, for both \( B < 2N - 1 \) and \( B \geq 2N - 1 \), the sum rate achieved by explicit constructions is strictly larger than that of separate encoding. The sum rate of separate encoding is derived though a fixed proportion of the two sources of symbols, which is equal to that in our approach. The sum rate of separate encoding is denoted as $R_{\text{sep}}$. And we can prove the following inequality holds:
\begin{equation}
\label{sep<}
\begin{aligned}
    R_{\text{sep}} < \left\{
    \begin{array}{ll}
        \displaystyle\frac{T_{\mathrm{v}} - 2N + 2}{T_{\mathrm{v}} - 2N + 2 + B}, & \text{for } B \geq 2N - 1; \\[10pt]
        \displaystyle\frac{T_{\mathrm{v}} - B + 1}{T_{\mathrm{v}} + 1}, & \text{for } B < 2N - 1.
    \end{array}
    \right.
\end{aligned}
\end{equation}

The detailed proof of this statement is provided in Appendix E.

To quantitatively illustrate the superiority in sum rate offered by our explicit constructions, we conducted an analysis with fixed channel parameters \( B = 9 \) and \( N = 3 \). Table I presents a detailed relative gain in sum rate for varying \( T_{\mathrm{v}} \) and \( T_{\mathrm{u}} \) compared to separate encoding. It is observed that when \( T_{\mathrm{v}} > T_{\mathrm{u}} + B \), there is a non-negligible gain in sum rate, reaching up to 16.00\%. These gain values indicate a significant improvement over separate encoding.

\begin{table}[h]
\centering
\caption{Gain values for varying \( T_{\mathrm{v}} \) and \( T_{\mathrm{u}} \) (in time steps)}
\label{table:gain_values}
\small 
\setlength{\tabcolsep}{4pt} 
\begin{tabular}{|c|c|c|c|c|c|c|}
\hline
\( T_{\mathrm{v}} / T_{\mathrm{u}} \) & 10     & 11     & 12     & 13     & 14     & 15     \\
\hline
20 & 16.00\% & 16.40\% & 16.67\% & 16.84\% & 16.93\% & 16.96\% \\
21 & 14.00\% & 14.57\% & 15.00\% & 15.31\% & 15.52\% & 15.65\% \\
22 & 12.00\% & 12.75\% & 13.33\% & 13.78\% & 14.11\% & 14.35\% \\
23 & 10.00\% & 10.93\% & 11.67\% & 12.24\% & 12.70\% & 13.04\% \\
24 & 8.00\%  & 9.11\%  & 10.00\% & 10.71\% & 11.29\% & 11.74\% \\
25 & 6.00\%  & 7.29\%  & 8.33\%  & 9.18\%  & 9.87\%  & 10.43\% \\
\hline
\end{tabular}
\end{table}

\section{Proof of Theorem 2}
\input{proof_achievable}

\section{A further discussion on the property of the proposed achievable region}
\input{Theorem1}

\section{Conclusion}
In this work, we investigated the scenario of transmitting two types of messages with distinct decoding deadlines through a channel that can introduce either one burst erasure of length at most \( B \) or \( N \) random erasures.
We proposed a non-trivial achievable region $\mathcal{R}$, which is strictly larger than the trivial region $\mathcal{B}$.
Subsequently, we presented a novel merging method for encoding the two streams, and additionally proposed explicit constructions based on this method.
These constructions contribute a non-trivial rate pair that significantly expands the achievable region. 
Further, through meticulous analysis, we demonstrated that the achievable region $\mathcal{R}$ encompasses the rate pairs of existing benchmark codes and aligns with the capacity region in the scenario where the sliding window model simplifies to the burst channel.
Finally, we explored the property of the proposed achievable region, affirming that the region $\mathcal{R}$ is the largest one under the merging method.

Future work could be extending the current study to more complicated channel models or adapt the merging strategy for more than two types of messages. This could be particularly relevant for IoT environments or complex multimedia streaming services. Furthermore, determining the capacity region for $(W,B,N)$-achievable block codes in sliding window model remains an open problem, offering a promising direction for future research.



\input{appendix}


\bibliographystyle{IEEEtran} 
\input{main.bbl}
\end{document}

%% file: construction_N.tex
Taking the merging size $m=B$, we define:
\begin{equation}
    \begin{gathered}
         k_{\mathrm{u}}=T_{\mathrm{u}}'-N+1,\\
       k_{\mathrm{v}}=T_{\mathrm{v}}-T_{\mathrm{u}}'+N-B,\\
         n=k_{\mathrm{u}}+k_{\mathrm{v}}+2B-m=T_{\mathrm{v}}+1.
    \end{gathered}
\end{equation}
Consequently, the obtained sum rate is given by:
\begin{equation}
R_{\mathrm{u}} + R_{\mathrm{v}} = \frac{T_{\mathrm{v}} - B +1}{T_{\mathrm{v}} + 1}.
\end{equation}

\begin{figure}[!b]
\centerline{\includegraphics[width=1\linewidth]{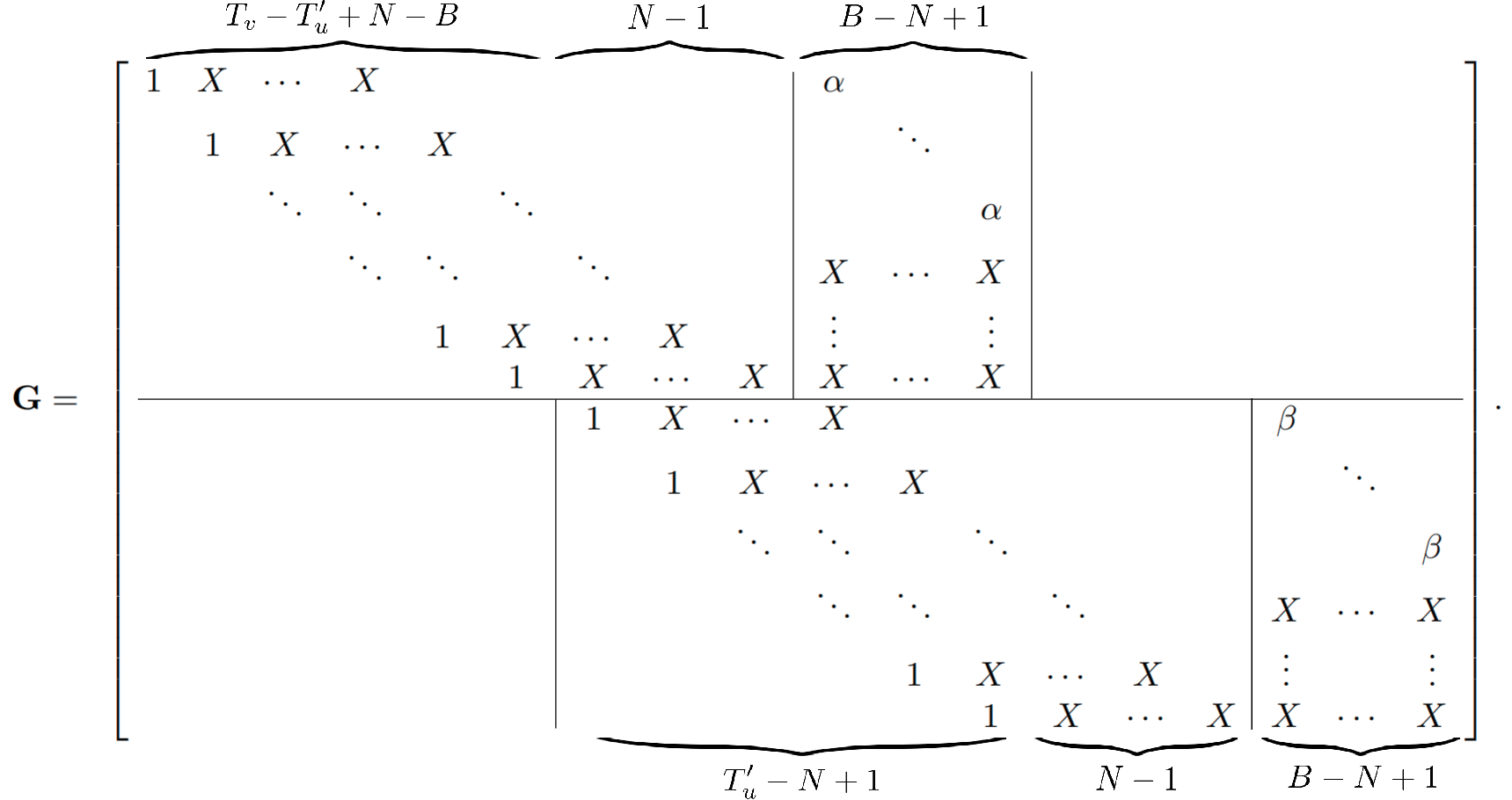}}
\caption{The generator matrix $\mathbf{G}$ of the proposed code when $B<2N-1$}
\label{Fig_cons_2}
\end{figure}

The codeword is as follows:
\begin{equation}
\label{cons_sizeN}
\begin{aligned}
&[x[0],x[1],\ldots,x[T_{\mathrm{v}}]] \\
=&[v[0],v[1],\ldots,v[T_{\mathrm{v}}-T_{\mathrm{u}}'+N-B-1], \\
&u[0],u[1],\ldots,u[T_{\mathrm{u}}'-N]] \mathbf{G},
\end{aligned}
\end{equation}
\begin{equation}
\left[x^{(\mathrm{v})}[0], x^{(\mathrm{v})}[1], \ldots, x^{(\mathrm{v})}\left[T_{\mathrm{v}}-T_{\mathrm{u}}'+N-1\right]\right]=\vec{v} G_1,
\end{equation}
\begin{equation}
\left[x^{(\mathrm{u})}[0], x^{(\mathrm{u})}[1] ,\ldots, x^{(\mathrm{u})}\left[T_{\mathrm{u}}'-N+B\right]\right]=\vec{u}G_2.
\end{equation}
where 
\begin{equation}
\mathbf{G}=\begin{array}{c@{\hspace{-5pt}}l}
\left[\begin{array}{ccc}
G_{1[1:k_{\mathrm{v}}]}&G_{1 [k_{\mathrm{v}}+1:k_{\mathrm{v}}+B]}&\mathbf{0}\\
\mathbf{0}&G_{2[1:B]}&G_{2[B+1:k_{\mathrm{u}}+B]}\
 \end{array}\right].
  \end{array}
  \end{equation}
Here, $G_1$ and $G_2$, with dimensions $k_{\mathrm{v}} \times (k_{\mathrm{v}} + B)$ and $k_{\mathrm{u}} \times (k_{\mathrm{u}} + B)$ respectively, are generator matrices of $(W, B, N)$-achievable rate-optimal block codes in Fig. \ref{Fig_geneM}. Additionally, Fig. \ref{Fig_cons_2} illustrates the explicit form of the generator matrix $\mathbf{G}$.

\begin{Remark}
Consistent with the case for \( B \geq 2N - 1 \), in the prescribed generator matrix $\mathbf{G}$,  \(\alpha\) is an element of the field \(\mathbb{F}_q\), while the element $\beta$ belongs to field $\mathbb{F}_{q^{2}} \backslash \mathbb{F}_{q}$.
\end{Remark}

The achievability of the constructed block code is established as follows:
\begin{Theorem}
\label{thm_Bsize_B<2N-1}
   For the case where $B < 2N - 1$,  the construction of the $(n,k_{\mathrm{v}},k_{\mathrm{u}},T_{\mathrm{v}},T_{\mathrm{u}})_{\mathbb{F}}$-block codes given in \eqref{cons_sizeN} is $(W,B,N)$-achievable.
\end{Theorem} 
\begin{proof}
    This construction primarily differs from the case when \( B \geq 2N-1 \) in terms of the parameters. The proof of its achievability mirrors that of Theorem 2, hence it is not reiterated here.
\end{proof}

%% file: proof_achievable.tex
Suppose that the position of the initial erasure is denoted as $i$, where $i$ is indexed starting from 0. We will demonstrate the achievability of the proposed construction for channel scenarios involving burst or random erasures in the channel. The explicit form of the generator matrix $\mathbf{G}$ can be found in Fig. 4, as presented in the above section.

-\textbf{Length-B burst} -

\textbf{Case 1:} $i \geq T_{\mathrm{v}}-T_{\mathrm{u}}'-N+1$.

It becomes evident that the information symbols at positions $i = 0, \ldots, T_{\mathrm{v}} - T_{\mathrm{u}}' - N$ are ascertainable. Notably, as the block matrix $G_{1[1:k_{\mathrm{v}}]}$ stands as an upper triangular matrix, the symbols $v[0], \ldots, v[k_{\mathrm{v}} - 1]$ can be decoded within their deadline $T_{\mathrm{v}}$.  Consequently, leveraging the generator matrix $G_1$, the symbols $\{x^{(\mathrm{v})}[T_{\mathrm{v}} - T_{\mathrm{u}}' - N + 1], \ldots, x^{(\mathrm{v})}[T_{\mathrm{v}} - T_{\mathrm{u}}' - N + B]\}$ become known.  By eliminating these quantities from the information symbols $\{x[T_{\mathrm{v}}-T_{\mathrm{u}}'-N+1],\ldots,x[T_{\mathrm{v}} - T_{\mathrm{u}}' - N + B]\}$, the problem transforms into the recovery of a length-$B$ burst within the urgent information symbols.  In virtue of the fact that generator matrix $G_2$ generates a $(k_{\mathrm{u}}+B,k_{\mathrm{u}},T_{\mathrm{u}}')_{\mathbb{F}}$-block code that is $(W, B, N)$-achievable, the urgent symbols can be decoded within delay $T_{\mathrm{u}}$. Therefore, all symbols can be successfully decoded within their deadlines.

\textbf{Case 2: $i < T_{\mathrm{v}}-T_{\mathrm{u}}'-N+1 $}.

To begin, we assert that $\{v[0], \ldots, v[i-1]\}$ can be decoded within their respective deadlines. Since the initial $i$ rows and columns of matrix $G_1$ form an upper triangular matrix. By employing $\{x[0], \ldots, x[i-1]\}$, we can successfully decode $\{v[0], \ldots, v[i-1]\}$.

Next, we will demonstrate that $v[i]$ can be decoded within the delay $T_{\mathrm{v}}$. Given that $i < T_{\mathrm{v}} - T_{\mathrm{u}}' - N + 1$, the number of burst erasures in $G_2$ must be less than $B$. According to the construction of $G_2$ as defined by equation \eqref{eq_urgent}, we can get $x^{(\mathrm{u})}[0]$ when we obtain $\{x^{(\mathrm{u})}[B],\ldots,x^{(\mathrm{u})}[T_{\mathrm{u}}']\}$. Similarly, we can get $x^{(\mathrm{u})}[1]$ when we obtain $\{x^{(\mathrm{u})}[B], \ldots, x^{(\mathrm{u})}[T_{\mathrm{u}}' + 1]\}$, and so forth. Notably, the first erased symbol is $x[i] = x^{(\mathrm{v})}[i]$. Therefore, we can obtain the set $\{x^{(\mathrm{v})}[\text{max}\{T_{\mathrm{v}}-T_{\mathrm{u}}'-N+1, i+B\}], \ldots, x^{(\mathrm{v})}[\text{min}\{T_{\mathrm{v}} - T_{\mathrm{u}}' - N + B, i+T_{\mathrm{v}}-T_{\mathrm{u}}'\}] \}$ by subtracting $\{x^{(\mathrm{u})}[\max\{0, i+B-k_{\mathrm{v}}\}], \ldots, (x^{(\mathrm{u})}[\min\{B - 1, i + N - 1\}]]$ from $\{x[\max\{T_{\mathrm{v}}' - T_{\mathrm{u}}' - N,i+B\}], \ldots, x[\min\{T_{\mathrm{v}} - T_{\mathrm{u}}' - N + B,T_{\mathrm{v}} - T_{\mathrm{u}}' + i\}]\}$.

Consequently, assume, without loss of generality, that $i\leq B-N$. In the case where $i > B - N$, the discussion is similar, but there is no need to account for the delay constraint to decode $v[i]$. As ${v[0], \ldots, v[i-1]}$ are already decoded by time $i+T_{\mathrm{v}} $, we eliminate the information symbols $\{v[0], \ldots, v[i-1]\}$ from $\{x^{(\mathrm{v})}[i + B], \ldots, x^{(\mathrm{v})}[i + T_{\mathrm{v}} - T_{\mathrm{u}}' - 1]\}$. Let $\{\widetilde{x}[i + B], \ldots, \widetilde{x}[i + T_{\mathrm{v}} - T_{\mathrm{u}}' - 1]\}$ denote the symbols that remain after this cancellation process.

Upon revisiting the second item in Lemma 2, it is noted that the block $\mathbf{H}_2$ in the matrix $G_1$, which is outlined with red dotted lines in Fig. 1, serves as the generator matrix of a $(k_{\mathrm{v}}-(B-N+1)+B, k_{\mathrm{v}}-(B-N+1))_{\mathbb{F}_q}$ MDS code, equivalently a $(T_{\mathrm{v}}-T_{\mathrm{u}}', T_{\mathrm{v}}-T_{\mathrm{u}}'-B)_{\mathbb{F}_q}$ MDS code. Furthermore, it is observed that the symbols $\{\widetilde{x}[i+B], \dots, \widetilde{x}[i+T_{\mathrm{v}}-T_{\mathrm{u}}'-1]\}$ are part of this $(T_{\mathrm{v}}-T_{\mathrm{u}}', T_{\mathrm{v}}-T_{\mathrm{u}}'-B)_{\mathbb{F}_q}$ MDS code. The count of these symbols, totaling $T_{\mathrm{v}}-T_{\mathrm{u}}'-B$, is adequate for the decoding of this MDS code. Consequently, this enables the retrieval of the symbols $\{v[B-N+1], \ldots, v[k_{\mathrm{v}}]\}$. Therefore, upon subtracting symbols $\{v[B - N + 1], \ldots, v[k_{\mathrm{v}}]\}$ from $x^{(\mathrm{v})}[i + T_{\mathrm{v}} - T_{\mathrm{u}}']$, we are able to recover $v[i]$. As a result, the overall time delay required to recover $v[i]$ does not exceed $T_{\mathrm{v}} - T_{\mathrm{u}}' + T_{\mathrm{u}}' = T_{\mathrm{v}}$. Employing a similar line of reasoning, all information symbols can be decoded within their respective deadlines.

-\textbf{N random erasures} -

\textbf{Case 1:} $i \geq T_{\mathrm{v}} - T_{\mathrm{u}}' - N + 1$.

Similarly, it becomes evident that the information symbols at positions $\{0, 1, \dots, T_{\mathrm{v}}' - T_{\mathrm{u}}' - N\}$ are known, enabling the decoding of $\{v[0], \dots, v[k_{\mathrm{v}} - 1]\}$. Consequently, by employing the generator matrix $G_1$, the values $\{x^{(\mathrm{v})}[T_{\mathrm{v}} - T_{\mathrm{u}}' - N], \ldots, x^{(\mathrm{v})}[T_{\mathrm{v}} - T_{\mathrm{u}}' - N + B]\}$ are determined. By eliminating these quantities from $\{x[T_{\mathrm{v}} - T_{\mathrm{u}}' - N], \ldots, x[T_{\mathrm{v}} - T_{\mathrm{u}}' - N + B]\}$, the problem simplifies to the recovery of $N$ random erasures within the urgent information symbols. Given that the generator matrix $G_2$ generates a $(k_{\mathrm{u}}+B,k_{\mathrm{u}},T_{\mathrm{u}}')_{\mathbb{F}}$-block code that is $(W, B, N)$-achievable, the erasures can be recovered, thus enabling the successful decoding of all information symbols within their respective deadlines.

\textbf{Case 2:} $i < T_{\mathrm{v}} - T_{\mathrm{u}}' - N + 1$.

Subsequently, we will address this scenario in two distinct subcases.
\begin{itemize}
    \item[] \textbf{Subcase 1:} symbol $x[T_{\mathrm{v}}]$ is erased.
    
    Obviously, the number of erasures within the time interval $[0:T_{\mathrm{v}} - 1]$ is less than or equal to $N - 1$. As per Lemma \ref{MDS_lemma}, the matrix $\widetilde{G}$ is a generator matrix of a $(T_{\mathrm{v}}-N+1,T_{\mathrm{v}}-2N+2)_{\mathbb{F}_q}$ MDS code. Thus, all information symbols can be decoded.

    \item[] \textbf{Subcase 2:} symbol $x[T_{\mathrm{v}}]$ is not erased.

    Symbols $\{x[0],\cdots,x[T_{\mathrm{v}}-1]\}$ have up to $N$ random erasures. As per Lemma \ref{MDS_lemma}, any $T_{\mathrm{v}} - 2N + 2$ rows of the sub-matrix $\widetilde{G}$, which is generated by the first $T_{\mathrm{v}}-N+1$ columns of $\mathbf{G}$, are linearly independent. Therefore, before the time $T_{\mathrm{v}}-1$, we can obtain $T_{\mathrm{v}}-2N+1$ linearly independent equations with respect to $T_{\mathrm{v}}-2N+2$ variables $\{v[0],\cdots,v[k_{\mathrm{v}}-1],u[0],\cdots,u[k_{\mathrm{u}}-1]\}$. Given that each element in $\widetilde{G}$ lies within the field $\mathbb{F}_{q}$, the Gaussian elimination process permits us to represent every symbol as a linear combination. It can be succinctly expressed as:
    \begin{align}
    \label{20}
    u[j]=\alpha_j + \beta_j u[0],\\
    \label{21}
    v[k]=\widetilde{\alpha}_k + \widetilde{\beta}_k u[0],
    \end{align}
    where $\alpha_j, \beta_j, \widetilde{\alpha}_k, \widetilde{\beta}_k \in \mathbb{F}_{q}$, $j\in \{0,1,\dots,T_{\mathrm{u}}'-N \}, k \in \{0,1,\dots,T_{\mathrm{v}}-T_{\mathrm{u}}'-N\}$.
    
    Consider the $T_{\mathrm{v}}+1$-th column of matrix $\mathbf{G}$, presenting the following equation:
    \begin{equation}
    \label{22}
        x[T_{\mathrm{v}}]=\beta u[0]+ \sum_{j=B-N}^{T_{\mathrm{u}}'-N}X_{j,T_{\mathrm{u}}'+N-1}u[j] ,
    \end{equation}
    where $X_{i,j}$ denotes element of the $i$-th row and $j$-th column in matrix $G_2$.
    
    Given that $\beta \in \mathbb{F}_{q^{2}} \backslash \mathbb{F}_{q}$, when substituting equation \eqref{20} into equation \eqref{22}, the coefficient preceding $u[0]$ in equation \eqref{22} is non-zero. As a result, we can decode $u[0]$ within the delay $T_{\mathrm{u}}'$. Utilizing equation \eqref{21}, we are able to decode $v[i]$. The cumulative delay for decoding $v[i]$ does not surpass $T_{\mathrm{v}} - T_{\mathrm{u}}' + T_{\mathrm{u}}' = T_{\mathrm{v}}$. Furthermore, all the information symbols can be decoded within their respective deadlines.
\end{itemize}

%% file: Theorem1.tex
In the above sections, we propose a novel merging method and explicit constructions of $(W,B,N)$-achievable streaming codes to attain a non-trivial corner point $(a,b)$, thus getting the achievable region $\mathcal{R}$ by a convex combination.
However, the direct selection of the corner point $(a,b)$ to construct the achievable region $\mathcal{R}$ by convex combination leads to a question: whether there exist other $(W,B,N)$-achievable streaming codes that might result in a larger achievable region defined by their corresponding rate pairs. 
It is challenging to identify the converse result for the general family of codes, which is stated as an open question in \cite{fong2020opmulti}.
Thus, it is natural to determine whether there exist $(W,B,N)$-achievable streaming codes based on the merging scheme we propose that could potentially lead to the creation of a larger achievable region.

In this section, we discuss the property of achievable region $\mathcal{R}$ by proposing two converse bounds specific to the merging method detailed in Section IV.
These two converse bounds prove that the achievable region $\mathcal{R}$ we present is the largest one attainable under the merging scheme.
Specifically, the property of the achievable region $\mathcal{R}$ is stated in the following theorem.
%
%

%

%
%
%
%
%
%

\begin{figure}[!t]
\centering
\includegraphics[width=1\linewidth]{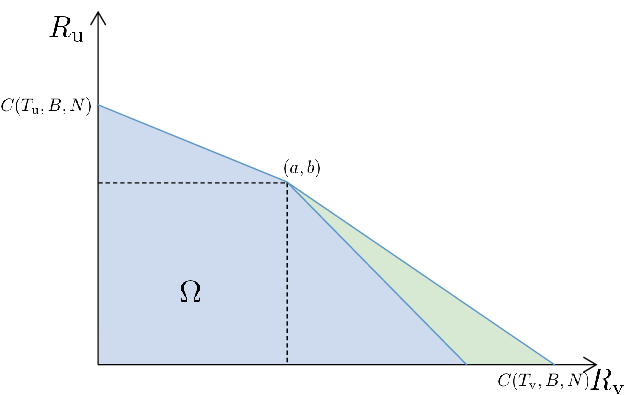}
\caption{The blue area represents a superset of the domain $\Omega$, while the entire area denotes the achievable $\mathcal{R}$. The green area represents a subset of $\mathcal{R}$ that extends beyond the superset of domain $\Omega$.}
\label{omega}
\end{figure}

\begin{Theorem}
\label{omega_in_R}
    Let $\Omega$ be the minimal domain containing all the rate pairs that could correspond to $(W,B,N)$-achievable streaming codes based on merging scheme. Then, 
\begin{equation}
    \Omega \in \left\{ (R_{\mathrm{v}},R_{\mathrm{u}}) \middle|
    \begin{aligned}
    & R_{\mathrm{u}} + R_{\mathrm{v}} \leq R_{\text{sum}}, \\
    & R_{\mathrm{v}} + \frac{1}{C(T_{\mathrm{u}},B,N)}R_{\mathrm{u}} \leq 1, \\
    & (R_{\mathrm{v}},R_{\mathrm{u}}) \in \mathbb{R}^2_{+},
    \end{aligned}
    \right\}
\end{equation}
where 
\begin{equation}
\begin{aligned}
     R_{\text{sum}}&=
    \begin{cases}
        \displaystyle \frac{T_{\mathrm{v}}-2N+2}{T_{\mathrm{v}}-2N+2+B}, & \text{if } B\geq 2N-1, \\
        \displaystyle \frac{T_{\mathrm{v}}-B+1}{T_{\mathrm{v}}+1}, & \text{if } B<2N-1.
    \end{cases}\\
\end{aligned}
\end{equation}
The domain $\Omega$ is represented in Fig. \ref{omega}. Therefore, $\Omega \in \mathcal{R}$.
\end{Theorem}
\begin{proof}
    In the following subsections, we derive two upper bound of the rate pairs achievable though our merging scheme, which proves the first part. Therefore, it is obvious that $\Omega \in \mathcal{R}$.
\end{proof}

\subsection{Trivial Bound Derived by Merging Method}
    Based on the merging method, the parameters of the $(n,k_{\mathrm{v}},k_{\mathrm{u}},T_{\mathrm{v}},T_{\mathrm{u}})_{\mathbb{F}}$-streaming code can be denoted as follows,
    \begin{equation}
        \begin{gathered}
    k_{\mathrm{v}} = T_{\mathrm{v}}' - N + 1, \\
    k_{\mathrm{u}} = T_{\mathrm{u}}' - N + 1, \\
    n = T_{\mathrm{v}}' + T_{\mathrm{u}}' - 2N + 2 + 2B - m.
\end{gathered}
    \end{equation}
    where $m$ denotes the merging size.
    
    Consequently, the corresponding rate pair can be denoted as
    \begin{equation}
      \begin{aligned}
      (R_{\mathrm{v}},R_{\mathrm{u}})=  &(\frac{T_{\mathrm{v}}'-N+1}{T_{\mathrm{v}}'+T_{\mathrm{u}}'-2N+2+2B-m},\\
            &\frac{T_{\mathrm{u}}'-N+1}{T_{\mathrm{v}}'+T_{\mathrm{u}}'-2N+2+2B-m}).
        \end{aligned}
    \end{equation}

    Thus, we can derive a family of linear constraints on the rate pair $(R_{\mathrm{v}},R_{\mathrm{u}})$ for every $T_{\mathrm{u}}'$:
    \begin{equation}
        R_{\mathrm{v}}+\frac{1}{C(T_{\mathrm{u}}',B,N)}R_{\mathrm{u}}\leq 1.
    \end{equation}
    Therefore, taking the closure of these inequalities for all $T_{\mathrm{u}}'$ we obtain the broadest possible region for $(R_{\mathrm{v}},R_{\mathrm{u}})$ that is achievable though merging method:
    \begin{equation}
        \label{bound1}
         R_{\mathrm{v}}+\frac{1}{C(T_{\mathrm{u}},B,N)}R_{\mathrm{u}}\leq 1.
    \end{equation}
\subsection{Sum Rate Bound}

\begin{Proposition}
\label{N_bound}
For \(1 \leq m \leq N-1\),  each \((W,B,N)\)-achievable streaming code based on the proposed merging method must satisfy the inequality \eqref{pro_1}.
\begin{figure*}[b]
\begin{equation}
\label{pro_1}
\begin{cases}
T_{\mathrm{v}}' + T_{\mathrm{u}}' - N + m \leq T_{\mathrm{v}}, & \text{if } B \geq 2N-1 \text{ or }(B < 2N-1 \text{ and } m \leq B-N+1) , \\
T_{\mathrm{v}}' + T_{\mathrm{u}}' - 2N + B +1\leq T_{\mathrm{v}}, & \text{if }B < 2N-1 \text{ and } m > B-N+1.
\end{cases}
\end{equation}
\end{figure*}

\end{Proposition}
\begin{proof}
$ $

\begin{enumerate}

    \item \textbf{Case 1:} $B\geq 2N-1$.
    
    In this scenario, we observe that $B - N\geq N-1 $. Assume the initial overlapped symbol positioned at $i$, where $i$ is indexed starting form 0. Given that the total count of less urgent symbols is $T_{\mathrm{v}}' - N + 1 + B$, it follows that:
    \begin{equation}
    m + T_{\mathrm{v}}' \leq T_{\mathrm{v}}' + N - 1 \leq T_{\mathrm{v}}' - N + 1 + B.
    \end{equation}
    This suggests at least $T_{\mathrm{v}}'$ symbols precede the first overlap, implying $i \geq T_{\mathrm{v}}'$. The relationship between $i$ and $m$ is thus:
    \begin{equation}
        \label{i_m}
        i+m=k_{\mathrm{v}}+B.
    \end{equation}

     We claim that
    \begin{equation}
    \label{51}
        T_{\mathrm{v}}'+T_{\mathrm{u}}'-N+m\leq T_{\mathrm{v}}.
    \end{equation}

    \eqref{51} serves as a necessary condition for $(W, B, N)$-achievable block codes derived by the merging method in this scenario. Conversely, suppose that, when $T_{\mathrm{v}}' + T_{\mathrm{u}}' - N + m > T_{\mathrm{v}}$, a $(W, B, N)$-achievable block code can be constructed from our merging method. We will present a counterexample in the following to demonstrate that this assumption is incorrect.
    
    Consider a burst of length $B$ that begins at position $i - T_{\mathrm{v}}'$. The symbols $\{x[0], \ldots, x[i - T_{\mathrm{v}}' - 1], x[i - T_{\mathrm{v}}' + B], \ldots, x[i - T_{\mathrm{v}}' + T_{\mathrm{v}}]\}$ in this $(W, B, N)$-achievable block code can decode $v[i - T_{\mathrm{v}}']$. This represents that the following system of linear equations  can be solved for the erased variable $v[i-T_{\mathrm{v}}']$. 
\begin{align}
    \label{eq_sys_lin}
    [x[0],&\dots,x[i-T_{\mathrm{v}}'-1]]= \nonumber \\
    &[\Vec{v},\Vec{u}]G_{[1:i-T_{\mathrm{v}}']}, \\
    [x[i-T_{\mathrm{v}}'&+B],\dots, x[i-T_{\mathrm{v}}'+T_{\mathrm{v}}]]= \nonumber \\
    &[\Vec{v},\Vec{u}]G_{[i-T_{\mathrm{v}}'+B:i-T_{\mathrm{v}}'+T_{\mathrm{v}}+1]}.
\end{align}

    Without loss of generality, consider $T_{\mathrm{v}}=T_{\mathrm{v}}'+T_{\mathrm{u}}'-N+m-1$. Referring to \eqref{G_merge}, we extract the initial $i-T_{\mathrm{v}}'+T_{\mathrm{v}}+1$ columns of $\mathbf{G}$, remove 
    $B$ columns within the interval $I=[i-T_{\mathrm{v}}':i-T_{\mathrm{v}}'+B-1]$, and generate a sub-matrix denoted as $\mathbf{M}$. Taking \eqref{i_m} into $i-T_{\mathrm{v}}'+T_{\mathrm{v}}+1$, we can derive that $i-T_{\mathrm{v}}'+T_{\mathrm{v}}+1-B=k_{\mathrm{u}}+k_{\mathrm{v}}-1$, which means the number of the columns we take is $k_{\mathrm{u}}+k_{\mathrm{v}}-1$. This extracted sub-matrix is respresented as:
    \begin{equation}
    \mathbf{M}=\begin{array}{c@{\hspace{-5pt}}l}
    \left[\begin{array}{ccc}
    G_{1[1:i]\backslash I}&G_{1 [i+1:k_{\mathrm{v}}+B]}&\mathbf{0}\\
    \mathbf{0}&G_{2[1:m]}&G_{2[m+1:m+k_{\mathrm{u}}-1]}\
    \end{array}\right].
     \end{array}
    \end{equation}

    Considering the matrix representation directly, the decodability of $v[i-T_{\mathrm{v}}']$ implies that the unit vector ${\eta}_{i-T_{\mathrm{v}}'+1}$ can be represented using the columns of $\mathbf{M}$. That is, there exists a vector $\mathbf{h}$ such that

    \begin{equation}
    \mathbf{M}\mathbf{h}=\mathbf{\eta}_{i-T_{\mathrm{v}}'+1},
    \end{equation}
    where $\eta_{i-T_{\mathrm{v}}'+1}$ is an unit vector in $\mathbb{F}^{k_{\mathrm{u}}+k_{\mathrm{v}}}$.
    
    From the decoding steps of $(W,B,N)$-achievable block code in single flow with a length-$B$ burst which begins at position $i-T_{\mathrm{v}}'$ \cite{domanovitz2021anexplicit}, there exists a vector $\hat{\mathbf{h}}$, such that 
    \begin{equation}
        \hat{G}_{1[i+1]}=\alpha \xi_{i-T_{\mathrm{v}}'+1} +\hat{G}_1 \hat{\mathbf{h}},
    \end{equation}
    where $\hat{G}_1\triangleq\begin{bmatrix}
G_{1[1:i]\backslash I}&G_{1 [i+1:k_{\mathrm{v}}+B]}        
    \end{bmatrix}$, and $\xi_{i-T_{\mathrm{v}}'+1}$ is an unit vector in $\mathbb{F}^{k_{\mathrm{v}}}$. In essence, the formula specifies that the symbol $v[i-T_{\mathrm{v}}]$ can be decoded using \( x[i] \) in conjunction with other symbols preceding \( x[i] \) in the sequence. For a more comprehensive discussion on this decoding mechanism, please refer to \cite{domanovitz2021anexplicit}.
    
    Considering the $i+1$-th column of matrix $\mathbf{M}$, we observe that
    \begin{equation}
    \label{53}
        \begin{aligned}
        &\mathbf{M}_{[i+1]}=\begin{bmatrix}
             \hat{G}_{1[i+1]} \\
             \mathbf{0}
    \end{bmatrix}+\eta_{k_{\mathrm{u}}+1},\\
         \eta_{k_{\mathrm{u}}+1}&= \mathbf{M}_{[i+1]}-\begin{bmatrix}
             \hat{G}_{1} \\
             \mathbf{0}
         \end{bmatrix}\hat{\mathbf{h}}-\alpha \eta_{i-T_{\mathrm{v}}'+1}.\\
                   &=\mathbf{M}_{[i+1]}-\mathbf{M}\begin{bmatrix} \hat{\mathbf{h}} \\ \mathbf{0} \end{bmatrix}-
        \alpha \mathbf{M}\mathbf{h}.
        \end{aligned}
    \end{equation}
Additionally, utilizing item 1 in Lemma 2 and block matrix rank properties, we can derive the following inequality of rank,
    \begin{equation}
         \label{rank}
    \begin{aligned}   
k_{\mathrm{u}}+k_{\mathrm{v}}-1\geq &  \text{rank}(\mathbf{M})\\
            \geq&\text{rank}(\begin{bmatrix}
                \hat{G_1} &\mathbf{0}\\
                \mathbf{0}&G_{2[m+1:m+k_{\mathrm{u}}-1]}
            \end{bmatrix}).\\
            =&k_{\mathrm{u}}+k_{\mathrm{v}}-1.
    \end{aligned}
    \end{equation}
From \eqref{53}, $\eta_{k_{\mathrm{u}}+1}$ can be expressed as a linear combination of the columns in $\mathbf{M}$, implying that
\begin{equation}
          \text{rank}(\begin{bmatrix}
            \mathbf{M}&\eta_{k_{\mathrm{u}}+1}        \end{bmatrix})=\text{rank}(\mathbf{M})=k_{\mathrm{u}}+k_{\mathrm{v}}-1.
\end{equation}
Similarly, we derive
\begin{equation}
\scalebox{0.88}{$
    \text{rank}\left(\begin{bmatrix}
        \mathbf{M}&\eta_{k_{\mathrm{u}}+1} 
    \end{bmatrix}\right)\geq\text{rank}\left(\begin{bmatrix}
        \hat{G_1} &\mathbf{0}&0\\
        \mathbf{0}&G_{2[m+1:m+k_{\mathrm{u}}-1]}&\zeta_{1}
    \end{bmatrix}\right),$}
\end{equation}

    where $\zeta_{1}$ is an unit vector in $\mathbb{F}^{k_{\mathrm{u}}}$.
    
    As per item 1 in Lemma 2, columns in $\hat{G}_2\triangleq \begin{bmatrix}
        G_{2[m+1:m+k_{\mathrm{u}}-1]}&\zeta_{1}
    \end{bmatrix}$ are all linearly independent, that is
    \begin{equation}
       \text{rank} (\hat{G}_2)=k_{\mathrm{u}}.
    \end{equation}

    Thus,
    \begin{equation}
    \begin{aligned}
                  \text{rank}(\begin{bmatrix}
            \mathbf{M}&\eta_{k_{\mathrm{u}}+1} \end{bmatrix})&\geq 
            \text{rank}(\hat{G_1})+\text{rank}(\hat{G}_2)\\
        &\geq k_{\mathrm{v}}+k_{\mathrm{u}},
    \end{aligned}
    \end{equation}
    which is contradictory with \eqref{rank}.

    \item \textbf{Case 2:} $B<2N-1$.
    
    We discuss this case in two subcases.
    
    --\textbf{Subcase 1:} $m\leq B-N+1$.
    
    Since $i+m=k_{\mathrm{v}}+B$, this subcase also implies that $i\geq T_{\mathrm{v}}'$. Thus, this case is exactly the same as case 1 stated.

    --\textbf{Subcase 2:} $m>B-N+1$.
    
    In this subcase, it is evident that $i<T_{\mathrm{v}}'$, and we will prove the proposition by a contradiction. If there is a block code derived by merging method is $(W,B,N)$-achievable when $T_{\mathrm{v}}'+T_{\mathrm{u}}'-2N+1+B> T_{\mathrm{v}} $. Suppose there is a burst beginning at position $0$. Then the following system of linear equation can be solved for the erased symbol $v[0]$:
    \begin{equation}
        [x[B],\dots,x[T_{\mathrm{v}}]]=[\Vec{v},\Vec{u}]G_{[1:T_{\mathrm{v}}+1]}.
    \end{equation}

    Without loss of generality, we assume $T_{\mathrm{v}}=T_{\mathrm{v}}'+T_{\mathrm{u}}'-2N+B$.
    Similarly, we construct the following sub-matrix $\mathbf{M}$.
    \begin{equation}
    \scalebox{0.95}{$
    \mathbf{M}=\begin{array}{c@{\hspace{-5pt}}l}
    \left[\begin{array}{ccc}
    G_{1[B+1:i]}&G_{1 [i+1:k_{\mathrm{v}}+B]}&\mathbf{0}\\
    \mathbf{0}&G_{2[1:m]}&G_{2[m+1:m+k_{\mathrm{u}}-1]}\
     \end{array}\right],
      \end{array}$}
      \end{equation}
     and the number of columns in $\mathbf{M}$ is $k_{\mathrm{v}}+k_{\mathrm{u}}-1$.
     There exists a vector $\mathbf{h}$, such that
    \begin{equation}
\mathbf{M}\mathbf{h}=\mathbf{\eta}_{1}.
    \end{equation}
     We can also derive an contradiction with the rank of $\mathbf{M}$. The remainder of the proof is identical to that of Case 1, and is therefore omitted here.
\end{enumerate}
    
\end{proof}

\begin{Proposition}
\label{B_bound}
For \(N-1 \leq m \leq B\),  each \((W,B,N)\)-achievable streaming code based on the proposed merging method  must satisfy the inequality 
\begin{equation}
\begin{cases}
T_{\mathrm{v}}' + T_{\mathrm{u}}' \leq T_{\mathrm{v}}, & \text{if } B \geq 2N - 1,  \\
T_{\mathrm{v}}' + T_{\mathrm{u}}'+B - 2N+1  \leq T_{\mathrm{v}}, & \text{if } B < 2N - 1.
\end{cases}
\end{equation}
\end{Proposition}
\begin{proof}
$ $
\begin{enumerate}
    \item \textbf{Case 1: $B\geq 2N-1$}.
    
    Initially, assume that the first overlapped symbol is positioned at \( i \), and a length-$B$ burst begins at position \( 0 \). For simplicity, we also assume \( i \leq T_{\mathrm{v}}' \). In fact, when $i>T_{\mathrm{v}}'$, we can discuss this scenario by assuming a burst beginning at position $i-T_{\mathrm{v}}'$. The subsequent discussion for $i > T_{\mathrm{v}}'$ will be identical to that for $i \leq T_{\mathrm{v}}'$. Thus, we discuss the scenario when \( i \leq T_{\mathrm{v}}' \).
    
    In this case, we assert that for any $(W, B, N)$-achievable block codes adhering to our strategy, the inequality $T_{\mathrm{v}}' + T_{\mathrm{u}}' \leq T_{\mathrm{v}}$ holds. If, contrary to this, when $T_{\mathrm{v}}' + T_{\mathrm{u}}' > T_{\mathrm{v}}$, there exists a block code based on merging method that can decode the erased symbols within their deadlines. Therefore, $\{x[B],\dots, x[T_{\mathrm{v}}]\}$ can decode $v[0]$. That is, the following system of linear equation can be solved for the erased variable $v[0]$.
    \begin{equation}
    \label{pro2_recover}
         [x[B],\dots,x[T_{\mathrm{v}}]]=[\Vec{v},\Vec{u}]G_{[1:T_{\mathrm{v}}+1]}.
    \end{equation}

    We assume, without loss of generality, that $T_{\mathrm{v}} = T_{\mathrm{v}}' + T_{\mathrm{u}}' - 1$.  Similarly, we extract the initial $T_{\mathrm{v}}+1$ columns of $\mathbf{G}$ and remove the initial $B$ columns. The extracted sub-matrix is denoted as:
    \begin{equation}
    \mathbf{M}=\begin{array}{c@{\hspace{-5pt}}l}
    \left[\begin{array}{ccc}
    G_{1[B+1:i]}&G_{1 [i+1:k_{\mathrm{v}}+B]}&\mathbf{0}\\
    \mathbf{0}&G_{2[1:m]}&G_{2[m+1:l]}\
     \end{array}\right],
      \end{array}
      \end{equation}
    where $l=m+T_{\mathrm{u}}'-1-B+N$, and the number of columns in $\mathbf{M}$ is $T_{\mathrm{v}}-B+1$.

    Similarly, referring to \eqref{pro2_recover}, there exists a vector $\mathbf{h} \in \mathbb{F}^{k_{\mathrm{u}}+k_{\mathrm{v}}}$ satisfying
\begin{equation}
\label{eq_condi}
\mathbf{M}\mathbf{h}=\eta_1.
\end{equation}
    where $\eta_1$ is the unit vector in $\mathbb{F}^{k_{\mathrm{u}}+k_{\mathrm{v}}}$.
Considering the $T_{\mathrm{v}}'-B+1$-th column of $\mathbf{M}$,
let \begin{equation}
    \mu \triangleq \mathbf{M}_{[T_{\mathrm{v}}'-B+1]}-\alpha \eta_1=
    [\underbrace{0, \dots, X,}_{k_{\mathrm{v}}}\underbrace{X, \dots  , 0}_{k_{\mathrm{u}}}]^T.
\end{equation}
Combining with \eqref{eq_condi}, we can observe that
\begin{equation}
\label{67}
    \begin{aligned}
      \alpha  \mathbf{M}\mathbf{h}=\mathbf{M}_{[T_{\mathrm{v}}'-B+1]}-\mu,\\
\mu=\mathbf{M}_{[T_{\mathrm{v}}'-B+1]}-\alpha \mathbf{M}\mathbf{h}.
      \end{aligned}
\end{equation}

It is evident that the number of columns in the matrix $G_{2[m+1:l]}$ is strictly less than $k_{\mathrm{u}}$. Utilizing Lemma 3, the matrix $G_{2[m+1:l]}$ possesses column full rank, which implies
 \begin{equation}
         \label{rank_2}
    \begin{aligned}   
T_{\mathrm{v}}-B+1\geq &  \text{rank}(\mathbf{M})\\
            \geq&\text{rank}(\begin{bmatrix}
                G_{1[B+1:k_{\mathrm{v}}+B]} &\mathbf{0}\\
                \mathbf{0}&G_{2[m+1:l]}
            \end{bmatrix}).\\
            =&\text{rank}(G_{1[B+1:k_{\mathrm{v}}+B]})+\text{rank}(G_{2[m+1:l]})\\
            =&k_{\mathrm{v}}+(l-m)\\
            =&T_{\mathrm{v}}-B+1.
    \end{aligned}
    \end{equation}
According to \eqref{67}, $\mu$ can be expressed as a linear combination of the columns of matrix $\mathbf{M}$. That is, 
 \begin{equation}
    \label{rank_2_2}
    \begin{aligned}
             \text{rank}(\begin{bmatrix}
            \mathbf{M}&\mu       \end{bmatrix})=   \text{rank}(\mathbf{M})=T_{\mathrm{v}}-B+1.
    \end{aligned}
    \end{equation}

Considering the structure of \(\mu\), it is evident that \(\mu\) is composed of elements from \(G_1\) and the \(m-B+N\)-th column of \(G_2\). Specifically, \(\mu\) can be represented as \(\mu = [\underbrace{0, \dots, X,}_{k_{\mathrm{v}}}\underbrace{G_{2[m+B-N]}^T}_{k_{\mathrm{u}}}]^T\).

Therefore,
\begin{equation}
\resizebox{0.9\columnwidth}{!}{%
$
\text{rank}\left(\begin{bmatrix}
            \mathbf{M}&\mu \end{bmatrix}\right) \geq \text{rank}\left(\begin{bmatrix}
                G_{1[B+1:k_{\mathrm{v}}+B]} & \mathbf{0} & 0 \\
                \mathbf{0} & G_{2[m+1:l]} & G_{2[m-B+N]}
            \end{bmatrix}\right).
$
}
\end{equation}

From the construction of $(k_{\mathrm{u}}, k_{\mathrm{u}} + B, T_{\mathrm{u}})_{\mathbb{F}}$ $(W, B, N)$-achievable block code, we deduce that $G_{2[m-B+N]}$ cannot be linearly represented by the columns of \( \mathbf{G}_{2[m+1:l]} \). This can be obtained similarly to the proof of Lemma 3, since the structure of \( \mathbf{G}_{2[m+1:l]} \) is such that the \( (m-B+N) \)-th row of the element in each column is zero, while in \( \mathbf{G}_{2[m-B+N]} \), the \( (m-B+N) \)-th element is clearly non-zero.

Thus, we can derive
\begin{equation}
\begin{aligned}
    \text{rank}\left(\begin{bmatrix}
G_{2[m+1:l]} & G_{2[m-B+N]}
\end{bmatrix}\right) &= (l-m)+1\\
&=T_{\mathrm{u}}'-B+N.
\end{aligned}
\end{equation}

Consequently, we can infer that
\begin{equation}
\begin{aligned}
&\text{rank}\left(\begin{bmatrix}
\mathbf{M} & \mu
\end{bmatrix}\right)\\
\geq& \text{rank}(G_{1[B+1:k_{\mathrm{v}}+B]}) + \text{rank}\left(
\scalebox{0.9}{$\begin{bmatrix}
G_{2[m+1:l]} & G_{2[m-B+N]}
\end{bmatrix}$}
\right) \\
=& k_{\mathrm{v}} + (T_{\mathrm{u}}' - B + N)\\
=& T_{\mathrm{v}} - B + 2,
\end{aligned}
\end{equation}
which contradicts the equation $\eqref{rank_2}$.

\item \textbf{Case 2:} $B<2N-1$.

Since $m\geq N-1$, it it obvious that $i\leq T_{\mathrm{v}}'$. Similarly, without loss of generality, set $T_{\mathrm{v}} = T_{\mathrm{v}}' +T_{\mathrm{u}}'+B-2N$. We can define the following matrix:
\begin{equation}
\resizebox{0.85\columnwidth}{!}{
$\mathbf{M} = 
\begin{array}{c@{\hspace{-5pt}}l}
\left[\begin{array}{ccc}
G_{1[B+1:h]} & G_{1 [i+1:k_{\mathrm{v}}+B]} & \mathbf{0}\\
\mathbf{0} & G_{2[1:m]} & G_{2[m+1:m+k_{\mathrm{u}}-1]}
\end{array}\right].
\end{array}
$}
\end{equation}

 Using a similar approach as in Case 1, we can still demonstrate a contradiction through the ranks of $\mathbf{M}$ and $\begin{bmatrix}
    \mathbf{M}&\mu
\end{bmatrix}$.

\end{enumerate}

\end{proof}

\begin{Theorem}
When $T_{\mathrm{v}} > T_{\mathrm{u}} + B > T_{\mathrm{u}} > B$, there exists an upper bound on the sum rate of the proposed merging method, which can be represented as follows:
\begin{equation}
\begin{aligned}
    R_{\mathrm{v}}+R_{\mathrm{u}}&\leq
    \begin{cases}
        \displaystyle \frac{T_{\mathrm{v}}-2N+2}{T_{\mathrm{v}}-2N+2+B}, & \text{if } B\geq 2N-1, \\
        \displaystyle \frac{T_{\mathrm{v}}-B+1}{T_{\mathrm{v}}+1}, & \text{if } B<2N-1.
    \end{cases}\\
\end{aligned}
\end{equation}
\end{Theorem}
\begin{proof}
Firstly, we assert that the merge length satisfies $m \leq B$. If we assume, to the contrary, that the merge length is $m > B$, then upon removing any contiguous set of B columns from the generator matrix $\mathbf{G}$, the number of rows in $\mathbf{G}$ becomes strictly less than the number of columns. Therefore, at least one source symbol becomes undecodable.

Consequently, we analyze the bound in two distinct cases, which can be intuitively deduced from Proposition \ref{N_bound} and Proposition \ref{B_bound}.
\begin{enumerate}
    \item  \textbf{Case 1: $1\leq m\leq N-1$}.
    
    First, let us consider the case when $B\geq2N-1$. Referring  to Proposition \ref{N_bound}, we derive
    \begin{equation}
        \begin{aligned}
            R_{\mathrm{v}}+R_{\mathrm{u}}&=\frac{k_{\mathrm{u}}+k_{\mathrm{v}}}{k_{\mathrm{u}}+k_{\mathrm{v}}+2B-m}\\
            &\leq\frac{T_{\mathrm{v}}-N+2-m}{T_{\mathrm{v}}-N+2-2m+2B}\\
            &=\frac{1}{2}+\frac{T_{\mathrm{v}}-\frac{1}{2}N+1-B}{T_{\mathrm{v}}-N+2-2m+2B}\\
            &\leq \frac{T_{\mathrm{v}}-2N+3}{T_{\mathrm{v}}-3N+4+2B}.
        \end{aligned}
    \end{equation}

    Similarly, when $B<2N-1$, we derive
    \begin{equation}
        R_{\mathrm{v}}+R_{\mathrm{u}}\leq \frac{T_{\mathrm{v}}-B+1}{T_{\mathrm{v}}+B-N}.
    \end{equation}
    
    \item \textbf{Case 2: $N\leq m\leq B$}.
    
        Referring  to Proposition \ref{B_bound} and similar to Case 1, we obtain
    \begin{equation}
    \begin{aligned}
        R_{\mathrm{v}} + R_{\mathrm{u}}&\leq
        \begin{cases}
            \displaystyle \frac{T_{\mathrm{v}}-2N+2}{T_{\mathrm{v}}-2N+2+B}, & \text{if } B\geq 2N-1, \\
            \displaystyle \frac{T_{\mathrm{v}}-B+1}{T_{\mathrm{v}}+1}, & \text{if } B<2N-1.
        \end{cases}
    \end{aligned}
    \end{equation}
\end{enumerate}
By straightforward calculations, we derive that
\scalebox{0.9}{\parbox{\linewidth}{
\begin{equation}
\begin{aligned}
    \begin{cases}
        \displaystyle \frac{T_{\mathrm{v}}-2N+2}{T_{\mathrm{v}}-2N+2+B}>\frac{T_{\mathrm{v}}-2N+3}{T_{\mathrm{v}}-3N+4+2B}, & \text{if } B\geq 2N-1, \\
        \displaystyle \frac{T_{\mathrm{v}}-B+1}{T_{\mathrm{v}}+1}\geq\frac{T_{\mathrm{v}}-B+1}{T_{\mathrm{v}}+B-N}, & \text{if } B<2N-1.
    \end{cases}
\end{aligned}
\end{equation}
}}

Hence, the proof is completed.
\end{proof}

%% file: appendix.tex
\section*{Appendix A \\Detailed Proof of Lemma 1} 
As shown in Table \ref{diag_table}, the symbols highlighted in the same color diagonally (in the $\searrow$ direction) are encoded using the given $(W,B,N)$-achievable block code. That is,
\begin{equation*} 
\begin{aligned}
   \relax[x_i[0], x_{i+1}[1], \dots, x_{i+n-1}[n - 1]] \triangleq\\
   [s_i[0], s_{i+1}[1], \dots, s_{i+k-1}[k - 1]] \mathbf{G}.
\end{aligned}
\end{equation*}
Given the fact that the $(W,B,N)$-achievable block code corrects $B$-length burst or $N$ random erasures within the respective deadlines, symbols $\{s_i[0],\dots,s_i[k-1]\}$ can be decoded within the deadlines for any $(W,B,N)$-erasure sequence (cf. Definition 2).

As for the second part, we can verify that for any $m = 0,\dots,n-1$, we have
\begin{equation}
    \begin{aligned}
x_i[m]= & [s_{i-m}[0], s_{i-m+1}[1], \dots, s_{i-m+k-1}[k - 1]] \mathbf{G} \xi_{m+1}\\
=&\sum_{l=0}^{n-1}\mathbf{s}_{i-l}\mathbf{G}_l \xi_{m+1},
    \end{aligned}
\end{equation}
where $\xi_{m+1}= [\underbrace{0,\dots,0}_{m}, 1,0,\dots,0 ]^T$ denotes the $n\times 1$ unit vector. This completes the proof.

\begin{table}[ht]
\centering
\caption{A construction of applying diagonal interleaving to generate the $(W,B,N)$-achievable streaming code}
\label{diag_table}
\scalebox{0.9}{
\begin{tabular}{|c|c|c|c|c|c|}
\hline
$\mathbf{x}_{i-1}$ & $\mathbf{x}_{i}$ & $\mathbf{x}_{i+1}$ & $\mathbf{x}_{i+2}$ & $\dots$ & $\mathbf{x}_{i+n-1}$ \\ \hline
\textcolor{red}{$x_{i-1}[0]$} &\textcolor{blue}{$x_{i}[0]$} &  &  &  &   \\ \hline
 & \textcolor{red}{$x_{i}[1]$} &\textcolor{blue}{$x_{i+1}[1]$} &  & & \\ \hline
 &  & \textcolor{red}{$x_{i+1}[2]$}& \textcolor{blue}{$x_{i+2}[2]$} & & \\ \hline
&  & & \textcolor{red}{$\ddots$}& \textcolor{blue}{$\ddots$}&  \\ \hline
 & & & & \textcolor{red}{$x_{i+n-2}[n-1]$} & \textcolor{blue}{$x_{i+n-1}[n-1]$} \\\hline
\end{tabular}}
\end{table}

\section*{Appendix B \\detailed proof of Lemma 3} 

First, if a set of $k=T-N+1$ consecutive columns appears before the $(T+1)$-th column of the matrix $\mathbf{G}$, then by item 1 of Lemma 2, they are evidently linearly independent. Without loss of generality, let us assume there are $m$ $(m>0)$ columns before the $(T+1)$-th column and $n$ $(n>0)$ columns after the $T$-th column, with $m+n=k$. We extract these consecutive $k$ columns of matrix $\mathbf{G}$ to form a sub-matrix $\mathbf{M}$, which is given by:
\begin{equation}
\mathbf{M}=[\underbrace{\mathbf{G}{[i]},\dots,\mathbf{G}{[T]}}_{m},\underbrace{\mathbf{G}{[T+1]},\dots,\mathbf{G}{[i+k]}}_{n}],
\end{equation}
where $i$ is the index of the first column of the sub-matrix $\mathbf{M}$, and clearly, $i \geq N$ since $n>0$.

We will prove that the matrix $\mathbf{M}$ has full column rank. According to item 1 in Lemma 2, the first $m$ columns of $\mathbf{M}$ are linearly independent. Let us denote the set of these $m$ columns as $S_1$. Given that the block $\mathbf{H}_3$ in $\mathbf{G}$ from Fig. \ref{Fig_geneM} is a diagonal matrix, the last $n$ columns are evidently linearly independent. We denote the set of these $n$ columns as $S_2$.

Next, we demonstrate that each column in $S_1$ is independent of all columns in $S_2$. Examining the nonzero entries in block $\mathbf{H}_1$ from Fig. \ref{Fig_geneM} reveals that the first $i-N+1$ rows of entries in every column of $S_1$ are zero. Conversely, when observing the nonzero entries in block $\mathbf{H}_3$ from Fig. \ref{Fig_geneM}, it is evident that the first $i+k-(T+1)$, which is equal to $i-N$, rows of entries in every column of $S_1$ are nonzero.

Hence, each column in $S_1$ is linearly independent of the columns in $S_2$, which completes the proof.

\section*{Appendix C \\Detailed proof of Theorem 1} 
The achievability proof is in the next section, where we propose the explicit constructions for $(W,B,N)$-achievable block codes, whose rate pairs corresponds to $(a,b)$.  
    
Consequently, we claim that the corner point $(a,b)$ is strictly above the line between the extreme points $(C(T_{\mathrm{v}},B,N),0)$ and $(0,C(T_{\mathrm{u}},B,N))$. The line can be represented as:
\begin{equation}
\label{line_extreme}
    R_{\mathrm{u}}=-\frac{C(T_{\mathrm{u}},B,N)}{C(T_{\mathrm{v}},B,N)}R_{\mathrm{v}}+C(T_{\mathrm{u}},B,N).
\end{equation}
By substituting $a$ into equation \eqref{line_extreme}, it suffices to demonstrate that the value is strictly less than $b$. When $B\geq 2N-1$, we have
\begin{equation}
\resizebox{0.99\columnwidth}{!}{%
$
    \begin{aligned}
        &-\frac{C(T_{\mathrm{u}},B,N)}{C(T_{\mathrm{v}},B,N)}a+C(T_{\mathrm{u}},B,N)-b\\
        =&-\frac{C(T_{\mathrm{u}},B,N)}{C(T_{\mathrm{v}},B,N)}\frac{T_{\mathrm{v}}-T_{\mathrm{u}}+N+1}{T_{\mathrm{v}}+2N+2+B}+C(T_{\mathrm{u}},B,N)-\frac{T_{\mathrm{u}}-N+1}{T_{\mathrm{v}}-2N+2+B}\\
        =&C(T_{\mathrm{u}},B,N)(\frac{-1}{C(T_{\mathrm{v}},B,N)}\frac{T_{\mathrm{v}}-T_{\mathrm{u}}-N+1}{T_{\mathrm{v}}-2N+2+B}+1-\frac{T_{\mathrm{u}}-N+1+B}{T_{\mathrm{u}}-2N+2+B})\\
       <&C(T_{\mathrm{u}},B,N)(1-(\frac{T_{\mathrm{v}}-T_{\mathrm{u}}+N+1}{T_{\mathrm{v}}-2N+2+B}+\frac{T_{\mathrm{u}}-N+1+B}{T_{\mathrm{u}}-2N+2+B}))\\
       =&0
    \end{aligned}
$
}
\end{equation}
We can derive a similar inequality when $B< 2N -1$. Therefore, the proof is completed.

\section*{Appendix D \\ Detailed Proof of Lemma \ref{MDS_lemma}} 

$\widetilde{G}$ can be expressed in the form of a block matrix as shown in equation \eqref{Gpiao}.
 \begin{equation}
\resizebox{.99\hsize}{!}{
    \ensuremath{
 \label{Gpiao}
\widetilde{G}= \begin{bmatrix}
 G_{1[1:k_{\mathrm{v}}]}&G_{1[k_{\mathrm{v}}+1:k_{\mathrm{v}}+N-1]}&G_{1[k_{\mathrm{v}}+N:k_{\mathrm{v}}+B]}&\mathbf{0}\\
 \mathbf{0}&G_{2[1:N-1]}&G_{2[N:B]}&G_{2[B+1:k_{\mathrm{u}}+N-1]}
    \end{bmatrix}. }
    }
\end{equation}   

To establish that \(\widetilde{G}\) is an MDS code, we use the equivalent criterion of verifying that every \((k_{\mathrm{v}} + k_{\mathrm{u}}) \times (k_{\mathrm{v}} + k_{\mathrm{u}})\) submatrix of \(\widetilde{G}\) has full rank.
To prove this, we construct the following matrix \(\mathbf{H}\) by replacing the \(G_{1[k_{\mathrm{v}}+N:k_{\mathrm{v}}+B]}\) block in \(\widetilde{G}\) with a zero matrix. 
The matrix \( \mathbf{H} \) is given by:
\begin{equation}
\label{ap_lemma_H}
           \mathbf{H}= \begin{bmatrix}
 G_{1[1:k_{\mathrm{v}}]}&G_{1[k_{\mathrm{v}}+1,k_{\mathrm{v}}+N-1]}&\mathbf{0}\\
 \mathbf{0}&G_{2[1:N-1]}&G_{2[N:k_{\mathrm{u}}+N-1]}
    \end{bmatrix}.
\end{equation}
That is
\begin{equation}
\mathbf{H}=\left[\begin{array}{cccccccccc}
1 & X & \cdots & X & X & 0 & \cdots & 0 & \cdots & 0  \\
0 & 1 & X & \cdots & X & X & \cdots & 0 & \cdots & 0  \\
0 & 0 & \ddots & \ddots & \ddots & \ddots & \ddots & 0& \cdots &0 \\
\vdots & \vdots & & 1 & \ddots & \ddots & \ddots & \ddots  & & \vdots\\
\vdots & \vdots & & & \ddots & X & \cdots & X & X & 0  \\
0 & 0 & \cdots & \cdots & 0 & 1 & X & \cdots & X & X  
\end{array}\right].
\end{equation}

Given that \(\mathbf{H}\) is obtained by setting certain elements of the matrix \(\widetilde{G}\) to zero, it is evident that the rank of any \((k_{\mathrm{v}} + k_{\mathrm{u}}) \times (k_{\mathrm{v}} + k_{\mathrm{u}})\) submatrix in \(\widetilde{G}\) is at least as large as that of the corresponding submatrix in \(\mathbf{H}\). 
Thus, it suffices to show that every \((k_{\mathrm{v}} + k_{\mathrm{u}}) \times (k_{\mathrm{v}} + k_{\mathrm{u}})\) submatrix of \(\mathbf{H} \) is of full rank.
Consequently, we consider a \( (k_{\mathrm{v}} + k_{\mathrm{u}}) \times (k_{\mathrm{v}} + k_{\mathrm{u}}) \) submatrix of \( \mathbf{H} \), denoted as \( \hat{\mathbf{H}} \), structured as a block matrix:
\begin{equation}
   \hat{\mathbf{H}} = \begin{bmatrix}
 A & B & \mathbf{0} \\
 \mathbf{0} & C & D
 \end{bmatrix}.
\end{equation}

By analyzing the column composition of \( \hat{\mathbf{H}} \), we observe:
\begin{itemize}
    \item \( \begin{bmatrix} A^T & \mathbf{0} \end{bmatrix}^T \) comprises columns exclusive to block \( G_1 \), with column number \( a \leq k_{\mathrm{v}} \).
    \item \( \begin{bmatrix} \mathbf{0} & D^T \end{bmatrix}^T \) includes columns exclusive to block \( G_2 \), with column number \( c \leq k_{\mathrm{u}} \).
    \item \( \begin{bmatrix} B^T & C^T \end{bmatrix}^T \) contains columns with non-zero elements from both \( G_1 \) and \( G_2 \), with column number \( b \leq N-1 \).
\end{itemize}

It follows that \( a + b + c = k_{\mathrm{v}} + k_{\mathrm{u}} \), satisfying \( a + b \geq k_{\mathrm{v}} \) and \( b + c \geq k_{\mathrm{u}} \). Assuming \( a + b < k_{\mathrm{v}} \) leads to \( c > k_{\mathrm{u}} \), a contradiction, thus \( a + b \geq k_{\mathrm{v}} \).

Subdividing submatrices \( B \) and \( C \) of \( \hat{\mathbf{H}} \) into \( B = [B_1 \quad B_2] \) and \( C = [C_1 \quad C_2] \), with \( B_1, C_1 \) having \( b_1 = k_{\mathrm{v}} - a \) columns and \( B_2, C_2 \) having \( b_2 = k_{\mathrm{u}} - c \) columns, we have:
\begin{equation}
   \hat{\mathbf{H}} = \begin{bmatrix}
 A & B_1 & B_2 & \mathbf{0} \\
 \mathbf{0} & C_1 & C_2 & D
 \end{bmatrix}.
\end{equation}

It is obvious that 
\begin{equation}
    \text{rank}(\begin{bmatrix}
 A&B_1&B_2&\mathbf{0}\\
 \mathbf{0}&C_1&C_2&D
 \end{bmatrix})\geq \text{rank}(\begin{bmatrix}
 A&B_1&\mathbf{0}&\mathbf{0}\\
 \mathbf{0}&\mathbf{0}&C_2&D
 \end{bmatrix}).
\end{equation}
Employing item 1 in Lemma 2, block $[G_{1[1:k_{\mathrm{v}}]}\quad G_{1[k_{\mathrm{v}}+1,k_{\mathrm{v}}+N-1]}]$ is a generator matrix of a $(k_{\mathrm{v}}+N-1,k_{\mathrm{v}})_{\mathbb{F}_q}$ MDS code, and block $[G_{1[1:k_{\mathrm{u}}]}\quad G_{1[k_{\mathrm{u}}+1,k_{\mathrm{u}}+N-1]}]$ is a generator matrix of a $(k_{\mathrm{u}}+N-1,k_{\mathrm{u}})_{\mathbb{F}_q}$ MDS code. We establish that \( \text{rank}([A \quad B_1]) = k_{\mathrm{v}} \) and \( \text{rank}([C_2 \quad D]) = k_{\mathrm{u}} \).

Then, 
\begin{equation}
    \begin{aligned}
         k_{\mathrm{v}}+k_{\mathrm{u}}&\geq \text{rank}(\begin{bmatrix}
 A&B_1&B_2&\mathbf{0}\\
 \mathbf{0}&C_1&C_2&D
 \end{bmatrix})\\
 &\geq \text{rank}(\begin{bmatrix}
 A&B_1&\mathbf{0}&\mathbf{0}\\
 \mathbf{0}&\mathbf{0}&C_2&D
 \end{bmatrix})\geq k_{\mathrm{v}}+k_{\mathrm{u}},
    \end{aligned}
\end{equation}
confirming that \( \hat{\mathbf{H}} \) is of full rank. This completes the proof that \( \widetilde{G} \) is a $(T_{\mathrm{v}}-N+1,T_{\mathrm{v}}-2N+2)_{\mathbb{F}_q}$ MDS code. 

\section*{Appendix E \\ detailed proof of equation \eqref{sep<}} 
    \input{fraction}

%% file: fraction.tex
We illustrate the scenario where $B > 2N - 1$.

Given that
\begin{equation}
    \begin{aligned}
        k_{\mathrm{u}}=T_{\mathrm{u}}-N+1,\\
        k_{\mathrm{v}}=T_{\mathrm{v}}-N+1,
    \end{aligned}
\end{equation}
we adopt the same proportion of distinct information symbols for separate encoding. In the case of separate encoding, $k_{\mathrm{u}}$ urgent symbols will get encoded to $\frac{k_{\mathrm{u}}}{C(T_{\mathrm{u}},B,N)}$ symbols and similarly, $k_{\mathrm{v}}$ less urgent symbols will get encoded to $\frac{k_{\mathrm{v}}}{C(T_{\mathrm{v}},B,N)}$ symbols. Thus, the total symbols after encoding is up to $\frac{k_{\mathrm{u}}}{C(T_{\mathrm{u}},B,N)}+\frac{k_{\mathrm{v}}}{C(T_{\mathrm{v}},B,N)}$. Hence, the sum rate, denoted as $R_{\text{sep}}$ is given by:
\begin{equation}
   R_{\text{sep}}= \frac{k_{\mathrm{u}}+k_{\mathrm{v}}}{\frac{k_{\mathrm{u}}}{C(T_{\mathrm{u}},B,N)}+\frac{k_{\mathrm{v}}}{C(T_{\mathrm{v}},B,N)}}.
\end{equation}
That is,
\begin{equation}
\begin{aligned}
     R_{\text{sep}}&
     =\frac{T_{\mathrm{v}}-2N+2}{\frac{T_{\mathrm{u}}-N+1}{C(T_{\mathrm{u}},B,N)}+\frac{T_{\mathrm{v}}-T_{\mathrm{u}}-N+1}{C(T_{\mathrm{v}},B,N)}}\\
     &=\frac{T_{\mathrm{v}}-2N+2}{(T_{\mathrm{u}}-N+1+B)+(T_{\mathrm{v}}-T_{\mathrm{u}}-N+1)\frac{T_{\mathrm{v}}-N+1+B}{T_{\mathrm{v}}-N+1}}\\
     &<\frac{T_{\mathrm{v}}-2N+2}{(T_{\mathrm{u}}-N+1+B)+(T_{\mathrm{v}}-T_{\mathrm{u}}-N+1)}\\
     &=\frac{T_{\mathrm{v}}-2N+2}{T_{\mathrm{v}}-2N+2+B}.
\end{aligned}
\end{equation}

For the case where $B < 2N - 1$, a similar argument holds, leading to:
\begin{equation}
R_{\text{sep}} < \frac{T_{\mathrm{v}} - B + 1}{T_{\mathrm{v}} + 1}.
\end{equation}

Hence, we have successfully established the complete proof.